 \documentclass[preprint2,longabstract]{aastex}

\usepackage{aalongtable,lscape}

\newcommand{\vsini} {$v~\sin{i}$}
\newcommand{\kms} {km\,s$^{-1}$}
\newcommand{\Teff} {$T_{\rm eff}$}
\newcommand{\grav} {$\log{g}$}
\newcommand{\fastwind} {{\sc fastwind}}
\newcommand{\bonnsai} {{\sc bonnsai}}
\newcommand{\ioni}[2]{#1\,{\sc #2}}

\newcommand{\solu}[2]{#1\,$\pm$\,#2}

\shorttitle{Orbital and physical properties of $\sigma$~Ori~Aa,Ab,B}
\shortauthors{Sim\'on-D\'iaz et al.}

\begin{document}

%% LaTeX will automatically break titles if they run longer than
%% one line. However, you may use \\ to force a line break if
%% you desire.

\title{Orbital and physical properties of the $\sigma$~Ori~Aa,Ab,B triple system}

%% Use \author, \affil, and the \and command to format
%% author and affiliation information.
%% Note that \email has replaced the old \authoremail command
%% from AASTeX v4.0. You can use \email to mark an email address
%% anywhere in the paper, not just in the front matter.
%% As in the title, use \\ to force line breaks.

\author{S.~Sim\'on-D\'iaz}%\altaffilmark{1,2,3} and Ivan R. King\altaffilmark{1}}
\affil{Instituto de Astrof\'isica de Canarias, E-38200 La Laguna, Tenerife, Spain\\
Departamento de Astrof\'isica, Universidad de La Laguna, E-38205 La Laguna, Tenerife, Spain}
\email{ssimon@iac.es}

\author{J.~A.~Caballero}
\affil{Centro de Astrobiolog\'ia (CSIC-INTA), ESAC Campus, PO Box 78, E-28691 
Villanueva de la Ca\~nada, Madrid, Spain}

\author{J.~Lorenzo}
\affil{Departamento de F\'isica, Ingenier\'ia de Sistemas y Teor\'ia de la Se\~nal, Escuela 
Polit\'ecnica Superior, University of Alicante, Apdo.~99, E-03080 Alicante, Spain}

\author{J.~Ma\'iz~Apell\'aniz}
\affil{Instituto de Astrof\'isica de Andaluc\'ia (CSIC), Glorieta de la Astronom\'ia s/n, E-18008 Granada, Spain\\
Centro de Astrobiolog\'ia (CSIC-INTA), ESAC Campus, PO Box 78, E-28691 
Villanueva de la Ca\~nada, Madrid, Spain}

\author{F.~R.~N.~Schneider}
\affil{Argelander-Institut f\"ur Astronomie der Universit\"at Bonn, Auf dem H\"ugel 71, D-53121 Bonn, Germany
}

\author{I.~Negueruela}
\affil{Departamento de F\'isica, Ingenier\'ia de Sistemas y Teor\'ia de la Se\~nal, Escuela 
Polit\'ecnica Superior, University of Alicante, Apdo.~99, E-03080 Alicante, Spain}

\author{R.~H. Barb\'a}
\affil{Departamento de F\'isica, Universidad de La Serena, Benavente 980, La Serena, Chile
}

\author{R.~Dorda and A.~Marco}
\affil{Departamento de F\'isica, Ingenier\'ia de Sistemas y Teor\'ia de la Se\~nal, Escuela 
Polit\'ecnica Superior, University of Alicante, Apdo.~99, E-03080 Alicante, Spain}

\author{D.~Montes}
\affil{Departamento Astrof\'isica, Facultad de Ciencias F\'isicas, Universidad Complutense de Madrid, E-28040 Madrid, Spain
}

\author{A.~Pellerin}
\affil{Department of Physics \& Astronomy, State University of New York at Geneseo, 1 College Circle, Geneseo, NY 14454, USA.
}

\author{J.~Sanchez-Bermudez}
\affil{Instituto de Astrof\'isica de Andaluc\'ia (CSIC), Glorieta de la Astronom\'ia s/n, E-18008 Granada, Spain
}

\author{\'A.~S\'odor}
\affil{Konkoly Observatory, Research Centre for Astronomy and Earth Sciences, Hungarian Academy of Sciences, H-01121 Budapest, Hungary}

\and

\author{A.~Sota}
\affil{Instituto de Astrof\'isica de Andaluc\'ia (CSIC), Glorieta de la Astronom\'ia s/n, E-18008 Granada, Spain}

%% Notice that each of these authors has alternate affiliations, which
%% are identified by the \altaffilmark after each name.  Specify alternate
%% affiliation information with \altaffiltext, with one command per each
%% affiliation.

%\altaffiltext{1}{Visiting Astronomer, Cerro Tololo Inter-American Observatory.
%CTIO is operated by AURA, Inc.\ under contract to the National Science
%Foundation.}
%\altaffiltext{2}{Society of Fellows, Harvard University.}

%% Mark off your abstract in the ``abstract'' environment. In the manuscript
%% style, abstract will output a Received/Accepted line after the
%% title and affiliation information. No date will appear since the author
%% does not have this information. The dates will be filled in by the
%% editorial office after submission.

\begin{abstract}
We provide a complete characterization of the astrophysical properties of the $\sigma$~Ori~Aa,Ab,B hierarchical triple system, and an improved set of orbital parameters for the highly eccentric $\sigma$~Ori~Aa,Ab spectroscopic binary.
We compiled a spectroscopic dataset comprising 90 high-resolution spectra 
covering a total time span of 1963 days. 
We applied the Lehman-Filh\'es method for a detailed orbital analysis of the radial velocity curves and performed a combined quantitative spectroscopic analysis of the {$\sigma$~Ori~Aa,Ab,B} system by means of the stellar atmosphere code \fastwind. We used our own plus other available information on photometry and distance to the system for measuring the radii, luminosities, and spectroscopic masses of the three components. We also inferred evolutionary masses and stellar ages using the Bayesian code \bonnsai.
The orbital analysis of the new radial velocity curves led to a very accurate orbital solution of the $\sigma$~Ori~Aa,Ab pair. We provided indirect arguments indicating that $\sigma$~Ori~B is a fast rotating early-B dwarf. The \fastwind+\bonnsai\ analysis showed that the Aa,Ab pair contains the hottest and most massive components of the triple system while $\sigma$~Ori~B is a bit cooler and less massive. The derived stellar ages of the inner pair are intriguingly younger than the one widely accepted for the $\sigma$~Orionis cluster, at \solu{3}{1}~Ma.
The outcome of this study
will be of key importance for a precise determination of the distance to the $\sigma$~Orionis cluster, the interpretation of the strong X-ray emission detected for $\sigma$~Ori~Aa,Ab,B, and the investigation of the formation and evolution of multiple massive stellar systems and substellar objects.
\end{abstract}

%% Keywords should appear after the \end{abstract} command. The uncommented
%% example has been keyed in ApJ style. See the instructions to authors
%% for the journal to which you are submitting your paper to determine
%% what keyword punctuation is appropriate.

%\keywords{globular clusters: general --- globular clusters: individual(NGC 6397,
%NGC 6624, NGC 7078, Terzan 8}


\section{Introduction}
\label{section.introduction}

The fourth brightest star in the Orion's Belt is  \object{$\sigma$~Ori} ({\em Qu{\ae} ultimam baltei pr{\ae}cedit ad austrum}, 48~Ori, HD~37648).
This Trapezium-like system is also the brightest source of the relatively nearby, almost extinction-free, $\sim$3\,Ma-old \object{$\sigma$~Orionis} open cluster in the Ori~OB1b association, which is widely acknowledged as a cornerstone for the study of the stellar and substellar formation \citep{Gar67, Wol96, Bej99, Wal08, Cab08b, Cab13}.
Remarkably, the components in the eponymous $\sigma$~Ori stellar system illuminate and shape the celebrated, conspicuous \object{Horsehead Nebula} \citep{Pet05, Hab05, Goi06, Goi09, Com07, Rim12}, and a close photo-eroded pair of a very low-mass star and a brown-dwarf proplyd \citep{van03, San04, Cab05, Bou09, Hod09}.
Therefore, the amount of high-energy photons injected by the high-mass $\sigma$~Ori stars into the intra-cluster medium is not only a compulsory input for testing certain models of low-mass star and brown-dwarf formation \citep[e.g.,][]{Whi07}, but also for understanding the astrochemistry in the Horsehead photo-dissociation region \citep[][and references above]{Abe03, Pou03, War06, Bow09}.

Currently, we know that there are five stars with spectral types earlier than {B3} (i.e., massive stars) in the central arcminute of the cluster, including $\sigma$~Ori~Aa, Ab, B, D, and E \citep{Cab14}.
Contrarily to $\sigma$~Ori~D, which is a normal B2\,V star with large projected rotational velocity \citep[\vsini\,=\,200~\kms; ][]{Sim14}, the other four stars have extensively pulled the attention of the stellar community. 
In particular, $\sigma$~Ori~E is a famous helium-rich, magnetically-strong, peculiar, variable star with spectral type B2\,Vp \citep{Wal74, Lan78, Gro97, Tow13}, while $\sigma$~Ori~Aa,Ab,B is a very high-mass, hierarchical triple system.
It is made of a $\sim$0.25\,arcsec-wide astrometric binary, `Aa,Ab--B', which has not completed a full revolution yet since its discovery \citep[$P_{\rm astrom}$ = 156.7$\pm$3.0\,a --][]{Bur1892, Tur08, Cab08a, Cab14}, and a spectroscopic binary, `Aa--Ab', in a highly eccentric orbit with a period 400 times smaller than the astrometric period \citep[$P_{\rm SB2}$ = 143.5$\pm$0.5\,d --][]{Fro04, Mic50, Bol74, Mor91, Sti01, Sim11a}.

This paper aims to be a continuation of the study started by \citet[][Paper~I]{Sim11a}, who compiled 23 FIES spectra between Nov.~2008 and Apr.~2011, confirmed the presence of a third massive star component in the $\sigma$~Ori~AB system, and determined for the first time the orbital parameters of the Aa,Ab pair. Here we present the analysis of a much extended dataset comprising a total of 90 high resolution spectra spanning $\sim$\,5 years. Besides the refinement of the orbital parameters determined from a longer and much better sampled radial velocity curve, we perform a quantitative spectroscopic analysis of the combined spectra of $\sigma$~Ori~Aa,Ab,B. The derived spectroscopic parameters (effective temperatures, gravities, and projected rotational velocities) are complemented with photometric information on the system to provide the complete set of stellar astrophysical parameters, namely radii, luminosities, spectroscopic and evolutionary masses, plus estimates of the stellar ages and number of 
ionizing 
photons. The paper, which partially benefits from preliminary results from recent interferometric observations by \citet{Sch13} and \citet{Hum13}, concludes with a discussion on the global orbital and physical properties of the $\sigma$~Ori~Aa,Ab,B system, highlighting the importance of the results of our study for the interpretation of the strong X-ray emission of the triple system, a precise determination of the distance to the $\sigma$~Orionis cluster, and new observational clues in the investigation of the formation and evolution of stars and brown dwarfs at all mass domains.

%%% <<< ZVEZDA

%================================================================================
\section{Observations}\label{section.obs}
%================================================================================

\begin{figure}[t!]
   \centering
   \includegraphics[width=0.45\textwidth]{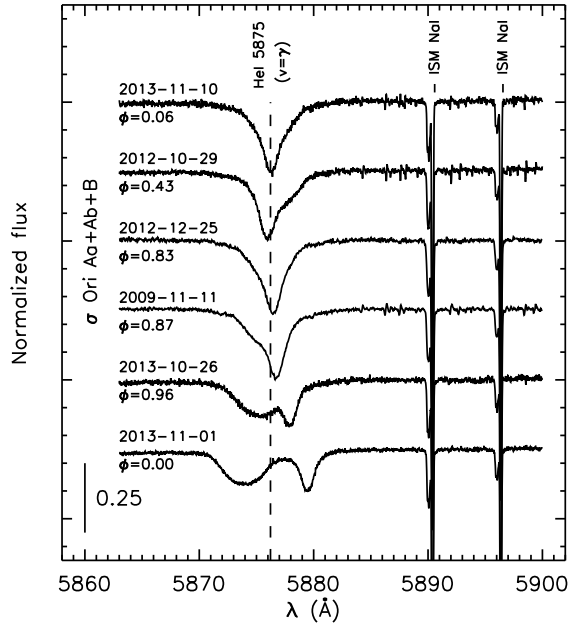}
      \caption{Spectra of $\sigma$~Ori~Aa,Ab,B in the region of the stellar \ioni{He}{i}\,$~\lambda$5875\,\AA\ and 
interstellar \ioni{Na}{i} lines at seven representative epochs.
The vertical dashed line indicates the wavelength of the \ioni{He}{i} line shifted to the systemic velocity ($\gamma$\,=\,+31.10\,\kms).
All spectra were corrected from telluric lines.}
         \label{figure.spectra}
   \end{figure}

After publication of Paper~I, we continued gathering high-resolution optical spectra with five different instruments, mainly attached to medium-sized telescopes. Our final sample of 90 spectra covers a total time-span of 1963 days, equivalent to almost 14 orbital periods of the Aa,Ab system. In particular, we could obtain over one third of the spectra near the periastron passages of Sep. 2010, Nov. 2011, Apr. 2012, Oct./Nov. 2013, and Mar. 2014.
\begin{itemize}
 \item Fifteen epochs were obtained with the FIES spectrograph ($R=46\,000$) at the 2.5~m Nordic Optical Telescope at El Roque de los Muchachos Observatory as part of the IACOB project \citep{Sim11b, Sim11d}. Together with the spectra used in Paper~I, these represent a total of 38 epochs.
 \item Thirty epochs were obtained with the HERMES spectrograph ($R=85\,000$) at the 1.2~m MERCATOR telescope at El Roque de los Muchachos Observatory, also as part of the IACOB project.
 \item Eleven epochs were obtained with the CAF\'E spectrograph ($R=65\,000$) at the 2.2~m telescope at Calar Alto Observatory as part of the CAF\'E-BEANS project (Negueruela et~al. 2014).
\item Six epochs were obtained with the HRS spectrograph ($R=30\,000$) at the 9.2~m Hobby-Eberly Telescope as part of the NoMaDS project \citep{Mai12}.
\item In addition, we included in our dataset four spectra obtained with the FEROS spectrograph ($R=48\,000$) at the 2.2~m telescope at La Silla Observatory as part of the OWN survey \citep{Bar10}.
\end{itemize}
Full descriptions of the various used instruments can be found in \citet[][FIES]{Tel14},
\citet[][HERMES]{Ras11}, \citet[][CAF\'E]{Ace13}, \citet[][HRS]{Tul98}, and \citet[][FEROS]{Kau99}.
The log of the observations is presented in Table \ref{table.rvs}, while Fig.~\ref{figure.spectra} shows portions of $\sigma$~Ori~Aa,Ab,B spectra at seven representative epochs. 
The typical signal-to-noise ratio (S/N) of all the spectra was above 200, which could be reached with 
exposure times of less than 10\,min in all telescopes. In the case of the two HERMES spectra obtained on 2013 Oct. 31 
(on the exact date of the closest quadrature to periastron passage), we increased the exposure time to reach a S/N\,$\sim$\,350.

The FIES, HERMES, and FEROS spectra were reduced using the corresponding available pipelines
(FIEStool\footnote{\tt http://www.not.iac.es/instruments/fies/fiestool/\ FIEStool.html},  %%% +++ ZVEZDA
HermesDRS\footnote{\tt http://www.mercator.iac.es/instruments/hermes/\ hermesdrs.php}, and  %%% +++ ZVEZDA
FEROS-DRS\footnote{\tt http://www.eso.org/sci/facilities/lasilla/\ instruments/feros/tools/DRS.html},  %%% +++ ZVEZDA
respectively). We used our own pipelines for reducing the CAF\'E and HET spectra. In all 
cases we used the information provided in the corresponding headers of the fits files to 
correct the spectra for heliocentric velocity, and our own routines implemented 
in IDL for continuum normalization.

We also obtained lucky imaging of the $\sigma$~Ori~Aa, Ab, B\, system with the AstraLux instrument at the Calar Alto 2.2~m telescope \citep{Hor08}. 
We used three different filters (Sloan $i$ and $z$ and a narrow-band filter centered at 9137 \AA) at five different epochs (from Jan. 2008 to Sep. 2013; see Table~\ref{AstraLuxtable}). 
The seeing varied between 0.9 and 1.3\,arcsec, but component B was clearly separated from A in the processed data in all cases (Aa and Ab components cannot be separated with lucky imaging).
The first epoch was discussed by \citet{Mai10}, where the reader is referred for further details on the data. 

%-------------------------------------------------------------
% Fig 1
%-------------------------------------------------------------
%-------------------------------------------------------------
%

%================================================================================
\section{Analysis and results}\label{section.analysis}
%================================================================================

\subsection{Radial velocity}\label{section.rv}

%-------------------------------------------------------------
% Fig 2
%-------------------------------------------------------------
   \begin{figure}[t!]
   \centering
   \includegraphics[width=0.45\textwidth]{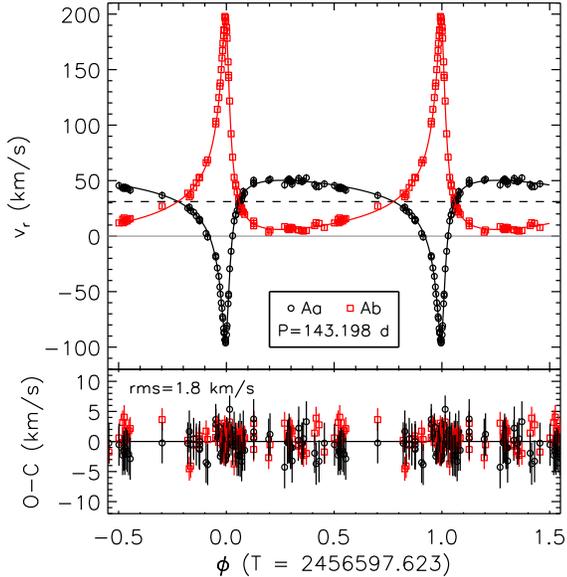}
      \caption{{\em Upper panel}: radial velocity curves of $\sigma$~Ori Aa (black circles) and Ab
      (red squares) phased to the period $P$ = 143.198\,d. 
      Error bar sizes are smaller than those of symbol size.
      The horizontal dashed line indicates the systemic velocity (see Table \ref{table.orbparameters}). 
      {\em Lower panel}: velocity residuals to the adopted fit for the two components.
      The root-mean-square of the fit is 1.8\,\kms.}
         \label{figure.rv}
   \end{figure}
%-------------------------------------------------------------

We followed a similar strategy to that described in Paper~I to revisit the orbital parameters of the Aa,Ab system resulting from the analysis of the extended radial velocity curves. 
In brief, we first used the \ioni{He}{i}\,$\lambda$5875\,\AA\ line to determine the radial velocity of the two components. 
To this aim, we performed a two-parameter cross correlation of the observed spectra with a grid of synthetic spectra built with two rotationally-broadened, radial velocity-shifted, \fastwind\ \ioni{He}{i}-line profiles. 
We used \vsini=135 and 35\,\kms\ for the Aa and Ab components, respectively \citep{Sti01,Sim11a}. The measured radial velocities, along with their corresponding uncertainties associated to the cross-correlations, are given in Table~\ref{table.rvs}. 

Next, we applied the Lehmann-Filh\'es method implemented in SBOP\footnote{\tt http://mintaka.sdsu.edu/faculty/etzel/}
(Etzel 2004) for a detailed orbital analysis of the radial velocity curves. We assumed as initial parameters those values indicated in Table 2 of Paper~I. Contrarily to our previous analysis, now we considered the period 
as a free parameter to be determined by SBOP. The resulting radial velocity curves, phased to the derived
period, are shown in Fig.~\ref{figure.rv}, and the revised orbital parameters are provided in
Table~\ref{table.orbparameters}.

The agreement with the solution presented in Paper~I is very good (but note the 180-degree indetermination of $\omega$ in Paper~I). 
However, as a consequence of the larger time-span and the better phase coverage (especially
around periastron passage) of the new radial velocity curves, 
we have improved in one order of magnitude the accuracy of all the resulting quantities. 
In particular, the period has been refined from 143.5$\pm$0.5\,d to 143.198$\pm$0.005\,d.
The new, more accurate, systemic velocity of the $\sigma$~Ori~Aa,Ab system (+31.10\,$\pm$\,0.16\,\kms) is also in very good agreement with the systemic velocity of the cluster as determined by \citet{Sac08} from single low-mass stars in $\sigma$~Orionis (+30.93\,$\pm$\,0.92\,\kms).
The mass ratio of the two components has increased from 1.23\,$\pm$\,0.07 to 1.325\,$\pm$\,0.006. 
The projected semimajor axes and masses have also been slightly modified.  %%% +++ ZVEZDA

%------------------------------------------------------
% Table 1
%------------------------------------------------------
\begin{table}[!t]
   \centering
      \caption[]{Revised orbital parameters of the $\sigma$~Ori~Aa,Ab system.} 
         \label{table.orbparameters}
     $$ 
         \begin{tabular}{l c c c}
            \hline
            \hline
            \noalign{\smallskip}
Parameter		&  \multicolumn{2}{c}{Value}				& Unit		\\
            \noalign{\smallskip}
            \hline
            \noalign{\smallskip}
$P_{\rm Aa,Ab}$		&  \multicolumn{2}{c}{143.198 $\pm$ 0.005 }		& d		\\
$T$			&  \multicolumn{2}{c}{2456597.623 $\pm$ 0.024} 	& d		\\
$e$			&  \multicolumn{2}{c}{0.7782 $\pm$ 0.0011}		& 		\\
$\gamma$		&  \multicolumn{2}{c}{+31.10 $\pm$ 0.16} 		& km\,s$^{-1}$	\\
$\omega$		& \multicolumn{2}{c}{199.98 $\pm$ 0.24}		& deg		\\
$M_{\rm Aa}/M_{\rm Ab}$	&  \multicolumn{2}{c}{1.325 $\pm$ 0.006} 		& 		\\
            \noalign{\smallskip}
            \hline
            \noalign{\smallskip}
				& Aa 			& Ab 					& 			\\
            \noalign{\smallskip}
            \cline{2-3}
            \noalign{\smallskip}
$K$			& 71.9 $\pm$ 0.3 		& 95.2 $\pm$ 0.3 				& km\,s$^{-1}$	\\
$a \sin{i}$		& 127.7 $\pm$ 0.6 		& 169.2 $\pm$ 0.6 				& $R_{\odot}$	\\
$M \sin^3{i}$		& 9.78 $\pm$ 0.07 	        & 7.38 $\pm$ 0.05 			& $M_{\odot}$	\\
            \noalign{\smallskip}
            \hline
         \end{tabular}
     $$ 
   \end{table}

\subsection{Photometry}
\label{section.photometry}

The $V$ magnitude of $\sigma$~Ori, including the three components, is 3.80\,mag \citep{Joh66, Lee68, Vog76, Duc01}. 
However, the determination of radii, luminosities, and masses of each component in  Section~\ref{section.physical} requires
the individual absolute magnitudes in the $V$ band. 
Also, this information is of interest for double-checking the dilution factors considered in the combined spectroscopic analysis in Section~\ref{section.fastwind}. 
For obtaining the extinction-corrected~$V$-band  %%% +++ ZVEZDA
absolute magnitudes of the three components we also need information about the magnitude difference 
between the A,B and Aa,Ab pairs, along with the extinction and distance to the stars.

\subsubsection{Magnitude difference between $\sigma$~Ori~A and B}

The difference in magnitude between the A (actually Aa,Ab) and B components has been measured by several authors using different techniques. 
With {\em Hipparcos} data, \citet{Per97} tabulated a magnitude difference $\Delta H_p$ = 1.21$\pm$0.05\,mag.
Using adaptive optics, \citet{ten00} measured the magnitude difference in three different optical filters ($\Delta V$ = 1.24$\pm$0.10, $\Delta R$ = 1.34$\pm$0.13, and $\Delta I$ = 1.25$\pm$0.15\,mag). 
Later on, \citet{Hor01} provided four speckle $V$-band differential photometry measures for $\sigma$~Ori~A,B ranging from 1.05 to 1.44\,mag (and a mean value of 1.18$\pm$0.08\,mag). 
Finally, \citet{Hor04} obtained again new speckle observations more consistent with previous adaptive optics measurements ($\Delta m_{\rm 503\,nm}$ = 1.30$\pm$0.07, $\Delta m_{\rm 648\,nm}$ = 1.25$\pm$0.03, $\Delta m_{\rm 701\,nm}$ $\sim$ 1.26~mag).

\begin{figure}[t!]
   \centering
   \includegraphics[width=0.45\textwidth]{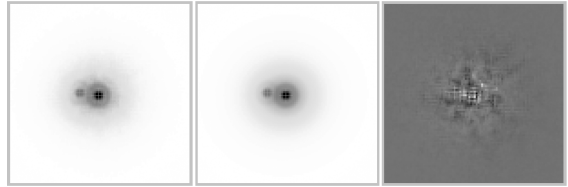}
      \caption{AstraLux observations of $\sigma$~Ori~A,B obtained on 2012-10-02 with the 9137\,\AA\ narrow-band filter. The three panels show the original data ({\em left}), PSF fit ({\em center}), and fit residual ({\em right}). The intensity scale is linear in the three cases, with the range corresponding to 0\,\% to 80\,\% of the maximum data pixel in the first two panels and to $-$5\,\% to $+$5\,\% of the maximum data pixel in the third one. Each panel is about 2.5\,arcsec on each side. North is up, and East is left.}
         \label{AstraLuxfigure}
   \end{figure}

We measured magnitude differences, angular separations ($\rho$), and position angles ($\theta$) in our own lucky imaging data (Section~\ref{section.obs}).
The AstraLux images were processed using the strategy described in \citet{Mai10} but with one important difference: 
instead of using a two-dimensional Gaussian for the point spread function (PSF) core, we used an obstructed Airy pattern with the parameters of the Calar Alto 2.2\,m telescope convolved with a two-dimensional Gaussian. 
An example of the data and fits is shown in Fig.~\ref{AstraLuxfigure}, while Table~\ref{AstraLuxtable} provides our measurements.

The dispersion of all magnitude differences, from 5030 to 9137\,\AA, is probably a consequence of the difficulty to characterize the PSF of the images and the proximity of the A and B components \citep[as indicated in][]{Mai10} rather than a colour effect, which is not expected at such early spectral types.
We gave more weight to our and {\em Hipparcos}~measurements and considered $\Delta V_{\rm A,B}$ = 1.20$\pm$0.05\,mag. 

%-------------------------------------------------------------
% Fig 3
%-------------------------------------------------------------
%-------------------------------------------------------------

% -------------------------------------------------------
% Table 2
% -------------------------------------------------------
\begin{table*}[!t]
   \centering
\caption{Relative astrometry and photometry of $\sigma$~Ori~A,B from the AstraLux data.}
\label{AstraLuxtable}
%\begin{scriptsize}
\begin{tabular}{ccccc}
\hline
\hline
\noalign{\smallskip}
Date       & Filter   & $\rho$            & $\theta$     & $m_{\rm B} - m_{\rm A}$      \\
(yyyy-mm-dd)           &          & [mas]       & [deg]   & [mag]           \\
\noalign{\smallskip}
\hline
\noalign{\smallskip}
2008-01-17 & $z$      & 253.9$\pm$2.0 & 92.9$\pm$0.6 & 1.319$\pm$0.035 \\
2011-09-13 & $z$      & 254.4$\pm$1.1 & 85.6$\pm$0.3 & 1.183$\pm$0.019 \\
2012-10-02 & 9137 \AA & 257.3$\pm$1.3 & 82.6$\pm$0.3 & 1.195$\pm$0.025 \\
2012-10-03 & $i$      & 256.8$\pm$0.9 & 82.6$\pm$0.3 & 1.201$\pm$0.016 \\
2013-09-16 & $i$      & 252.6$\pm$1.0 & 80.1$\pm$0.4 & 1.163$\pm$0.016 \\
\noalign{\smallskip}
\hline
\end{tabular}
%\end{scriptsize}
\end{table*}

\subsubsection{Magnitude difference between $\sigma$~Ori~Aa and Ab}

The case of $\sigma$~Ori~Aa and Ab is more complex, mainly because the angular 
separation of these two stars is so small that only interferometric observations
provide enough spatial resolution to resolve both components.
As a consequence, the few estimations of the Aa,Ab magnitude difference
found in the literature come from indirect arguments based on the derived spectral 
types of the two components, which in many cases were erroneously associated
to the A and B components. In particular, from the inspection of one of his 
high resolution spectrograms showing separated lines, \citet{Bol74} indicated that 
the redshifted component (Ab at +169.6~\kms) appeared to be 0.5\,mag fainter than 
the blueshifted component (Aa at --88~\kms). \citet{Edw76} derived 
an independent visual magnitude difference, $\Delta V_{\rm Aa,Ab}$ = 0.78\,mag, 
after following a more quantitative strategy aimed at determining the spectral class of the 
individual components of visual binaries. 

Only recently, interferometric observations with MIRC and NPOI by \citet{Sch13} and \citet{Hum13} have been able to provide a direct measurement of the magnitude difference between Aa and Ab. 
\citet{Sch13} measured a flux ratio between the two components in the near-infrared $H$ band of 0.58, 
which translates into a difference in magnitude of 0.59\,mag in $H$, in good agreement with 
the value proposed by \citet{Bol74}, but incompatible with the value estimated by Edwards (1976).
All in all, we assumed $\Delta V_{\rm Aa,Ab}$ = 0.59$\pm$0.05\,mag.

\subsubsection{Extinction}

\begin{figure}[t!]
   \centering
   \includegraphics[angle=90, width=0.49\textwidth]{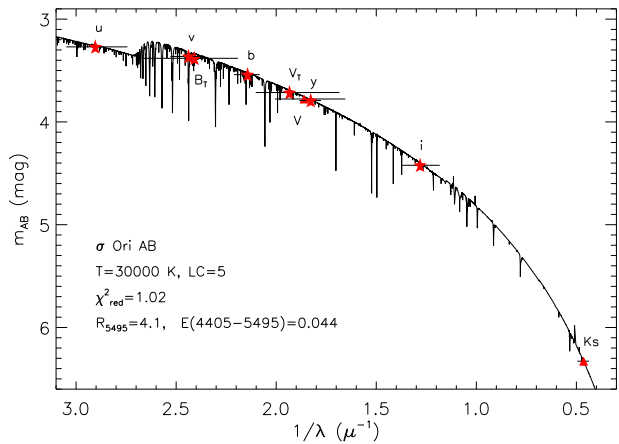}
\caption{{\sc chorizos} best spectral energy distribution for the $\sigma$~Ori~AB photometry. Red star symbols with error bars (horizontal for filter 
extent, vertical for uncertainty) indicate the used photometry (Tycho-2 $B_{T}V_{T}$, 2MASS $K_s$, Johnson-Cousins $VI$, 
and Str\"omgren $uvby$) in the AB system.}
\label{CHORIZOSfigure}
   \end{figure}

In most published photometry, $\sigma$~Ori~Aa,Ab,B appears as a single source. 
We collected from the literature Str\"omgren $uvby$, Tycho-2 $B_TV_T$, Johnson-Cousins $VI$, and 2MASS $K_s$ photometry for the system (2MASS $JH$ and {\em WISE} photometry saturated), and processed the data using the latest version of the photometric Bayesian code {\sc chorizos} \citep{Mai04}.
It incorporates the Milky Way spectral energy distribution grid of \cite{Mai13a}, the photometric calibration and zero points 
of \citet{Mai05a, Mai06, Mai07}, and the family of extinction laws of \citet{Mai13b} and \citet{Mai14}. 
We fixed the photometric luminosity class to 5.0 and we left four free parameters: 
$T_{\rm eff}$, 
$E(4405-5495)$ (amount of extinction), 
$R_{5495}$ (type of extinction), 
and $\log d$. 
We fitted nine photometric bands, and hence had five degrees of freedom. 
%%% ZVEZDA >>>
The resulting values for these four parameters and the best spectral energy distribution are shown in Table~\ref{CHORIZOStable} and Fig.~\ref{CHORIZOSfigure}, respectively.
The estimated extinction in the $V$ band is $A_{\rm V}$ = 0.18$\pm$0.02\,mag.
%%% <<< ZVEZDA

The {\sc chorizos} execution yielded a reduced $\chi^2$ value of 1.0, indicating that the photometry used is consistent and 
compatible with the used spectral energy distributions and extinction laws. The values of $E(4405-5495)$ and $A_{V}$ are compatible with 
previous results \citep{Lee68, Bro94, May08}. $R_{5495}$ shows a large uncertainty, as expected for an object with such a low extinction. The derived $T_{\rm eff}$ is consistent with a composite source made out of three bright stars, one hotter than 30\,kK, the another two cooler than that.
Finally, the derived distance ($d \approx$ 263\,pc) is considerably lower than other measurements, but this is just an artifact of the fitting and must not be used: 
since we are measuring the combined photometry of three stars and fixing the luminosity class, the resulting distance is artificially lower than the real one (see below). This problem with spectroscopic parallaxes is common for O-type systems due to the abundance of unresolved binaries.

%-------------------------------------------------------------
% Fig 4
%-------------------------------------------------------------
%-------------------------------------------------------------

% -------------------------------------------------------
% Table 3
% -------------------------------------------------------
\begin{table}[!t]
   \centering
\caption{{\sc chorizos} output parameters for $\sigma$~Ori.}
\label{CHORIZOStable}
\begin{tabular}{lcc}
\hline
\hline
\noalign{\smallskip}
Parameter             & Value & Unit \\
\noalign{\smallskip}
\hline
\noalign{\smallskip}
%$T_{\rm eff}$    & 29\,900$\pm$600     &   K       \\
$T_{*}$    & 29.9$\pm$0.6     &   kK       \\
$E(4405-5495)$ &   0.044$\pm$0.007   &   mag       \\
$R_{5495}$           &     4.1$\pm$0.8     &          \\
$\log d$         &   2.420$\pm$0.017   &         \\
$A_{V}$       &   0.18$\pm$0.02   &   mag       \\
%$V_{J,0}$       &   3.626$\pm$0.020   &   mag       \\
\noalign{\smallskip}
\hline
\end{tabular}
\end{table}

\subsubsection{Distance}
\label{sub.distance}

%%% ZVEZDA >>>

There have been different determinations of the distance to the $\sigma$~Orionis cluster, which have been derived and used in a variety of manners across the literature. 
In spite of the efforts put by several authors in obtaining an accurate and reliable value, there does not seem to be an acceptable consensus yet, with determinations ranging from 350 to 470\,pc.
\citet{Mai04+} gave a Lutz-Kelker-corrected value of $380^{+136}_{-87}$\,pc for $\sigma$~Ori based on the original {\it Hipparcos} measurement \citep{Per97}, but with a self-consistent (and non-constant) spatial distribution for the early-type stars in the solar neighborhood \citep{Lut73, Mai01, Mai05b}.
The new {\it Hipparcos} reduction of the raw data by \citet{van07} did not improve the original measurement.

From average {\em Hipparcos} parallax measurements of stars in Ori~OB1b, other authors have reported 
distances of up to 440\,pc \citep{Bro94, deZ99, Her05, CabDin08}. 
Alternative, non-parallactic determinations of the cluster distance have been mostly based on isochronal fitting.
However, perhaps because of the heterogeneity in used data and models, minimisation techniques, spectral 
types of stars, and subjective authors' assumptions, derived distances also range on a wide interval 
from 360$^{+70}_{-60}$ to 470$\pm$30\,pc \citep{She04, She08, Her05, Cab07, May08, Nay09}.

A classical distance-determination method, the dynamical parallax, was implemented in the cluster by
\citet{Cab08a}, who estimated $d \sim$ 385\,pc under the assumption that $\sigma$~Ori was {\em triple}. 
This estimation matches the recent determinations with the Navy Precision Optical Interferometer \citep[NPOI;][]{Hum13} and the Michigan Infra-Red Combiner for the CHARA Interferometer \citep[MIRC;][]{Sch13}.
From these interferometric data, \citet{Hum13} and \citet{Sch13} independently proposed a distance $d \sim$\,385\,pc.
Although optical interferometry will soon provide uncertainties of 1--2\,\% for the distance to $\sigma$~Ori 
(Schaefer et~al., in prep.), we assumed a more conservative error of 5\,\%, which translates into a heliocentric distance of $d$ = 385$\pm$19\,pc.

\subsection{The elusive $\sigma$~Ori~B component}

\citet{Bol74} was one of the first authors to postulate the presence of a third component
in the $\sigma$~Ori~A,B system \citep[although][had done it seven decades earlier]{Fro04}. 
He commented that the velocities and ``spectroscopic'' 
magnitude difference shown in one of his spectrograms indicated that none of the
spectroscopic components was the component B of the visual system and, furthermore, that there was no
evidence of B in his spectrograms. 
As shown below, our present-day, better quality, observational dataset confirms Bolton's statements. 

\begin{figure}[t!]
   \centering
   \includegraphics[width=0.46\textwidth]{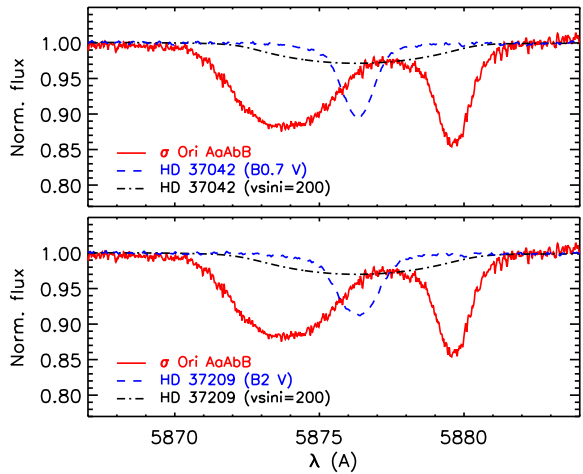}
      \caption{\ioni{He}{i}\,$\lambda$5875\,\AA\ line in the HERMES spectra of HD\,37\,042 (B0.7\,V, upper panel) 
      and HD\,37\,209 (B2\,V, lower panel) overplotted on the HERMES spectrum of $\sigma$~Ori~Aa,Ab,B obtained 
      on 2013-11-10 (red solid line). The spectra of the comparison stars have been shifted to the position
      where $\sigma$~Ori~B should be found ($V_r$ = $\gamma$ = \,31.10~\kms\ ), and diluted to account for the difference in 
      magnitude between the Aa, Ab and B components. Black dash-dotted lines: spectra of the comparison stars convolved
      to \vsini\,=\,200~\kms; blue dashed lines: original diluted spectra.}
         \label{figure.whereB}
   \end{figure}

Following the estimations by \citet{Har96}, component B should not be separated by more than $\sim$3\,\kms\ from the systemic velocity of the Aa,Ab pair. 
From the difference in magnitude relative to the Aa,Ab component, the B star should have an spectral 
type $\sim$B0--B2\,V. Thanks to the large radial velocities reached by $\sigma$~Ori~Aa and 
Ab during periastron passage (separated by up to 292\,\kms; see Table~\ref{table.orbparameters}), 
it might be possible to find spectroscopic features of the B component around phase zero. 
However, we did not find any sign of lines from an early B star close to the systemic velocity in any of the spectra obtained during periastron passage. 
This fact is illustrated by Fig.~\ref{figure.whereB}, where we show the spectrum of $\sigma$~Ori Aa,Ab,B with maximum separation between the lines of the Aa and Ab components together with high-S/N spectra of two well-investigated early B dwarfs.
The only possibility to reconcile photometry (i.e., the celebrated astrometric ``binary'' announced by Burnham in 1892) and spectroscopy (i.e., the absence of lines at $V_r \approx\gamma$) is to consider that $\sigma$~Ori~B has a large rotational velocity, of \vsini\,$\sim$\,200--300\,\kms. 

%-------------------------------------------------------------
% Fig 5
%-------------------------------------------------------------
%-------------------------------------------------------------

\subsection{Stellar parameters of $\sigma$~Ori~Aa, Ab, and B}
\label{section.parameters}

\subsubsection{Spectroscopic parameters}
\label{section.fastwind}

We used the HERMES spectrum taken on 2013 Oct. 11, the one with the largest
separation between the lines of the Aa and Ab components, to perform a {\em combined} quantitative 
spectroscopic analysis of the $\sigma$~Ori~Aa,Ab,B system. By {\em combined} we mean that the stellar parameters 
of the three components were obtained directly and simultaneously from the analysis of one of the 
original spectra. This type of analysis can be considered as opposite to the spectral 
disentangling option, in which the spectrum of each component is obtained and analyzed separately.
The combined synthetic spectra to be fitted to the observed one were constructed 
using spectra from the grid of \fastwind\ models with Solar metallicity \citep{Gon08, Sim10, Nie11} included in 
IACOB-GBAT \citep{Sim11c}. 
The spectra of each component was convolved to the corresponding \vsini, shifted in radial velocity, and scaled by a certain factor $d_i$, where $\sum d_i$=1. 
Then, the three synthetic spectra were added together, and the combined spectrum overplotted to the observed one. 

We fixed in our analysis the radial and projected rotational velocities of the three components. 
For components Aa and Ab we considered the values directly measured from the 2013-10-11 spectrum itself (see Table~\ref{t4}), while for component B we initially assumed a \vsini\,=\,200~\kms\ and a radial velocity equal to the systemic velocity of the Aa,Ab system (note the difference between the inclination $i$ of the rotation axis of each component, as quoted in \vsini, and the inclination of the orbit of the Aa,Ab system, as quoted in $M_{\rm dyn} \sin^3{i}$ in Table~\ref{table.orbparameters}). 
In view of the results from the spectroscopic analysis by \citet{Naj11}, we also fixed the associated helium abundances ($Y_{\rm He}$\,=\,0.10), microturbulence velocities ($\xi_{\rm t}$\,=\,5~\kms), and wind-strength parameters (log\,$Q$\,=\,--14.0) for the three stars. 
Last, we made use of the magnitude differences of previous sections to estimate the dilution factors. 
In particular, the Aa, Ab, and B components contributed to the global spectrum by 48\,\%, 28\,\%, and 24\,\%, respectively.
As a result, only the effective temperatures (\Teff) and gravities (\grav) remained as free parameters to be determined in the analysis. 

\begin{figure*}[t!]
   \centering
   \includegraphics[width=0.90\textwidth]{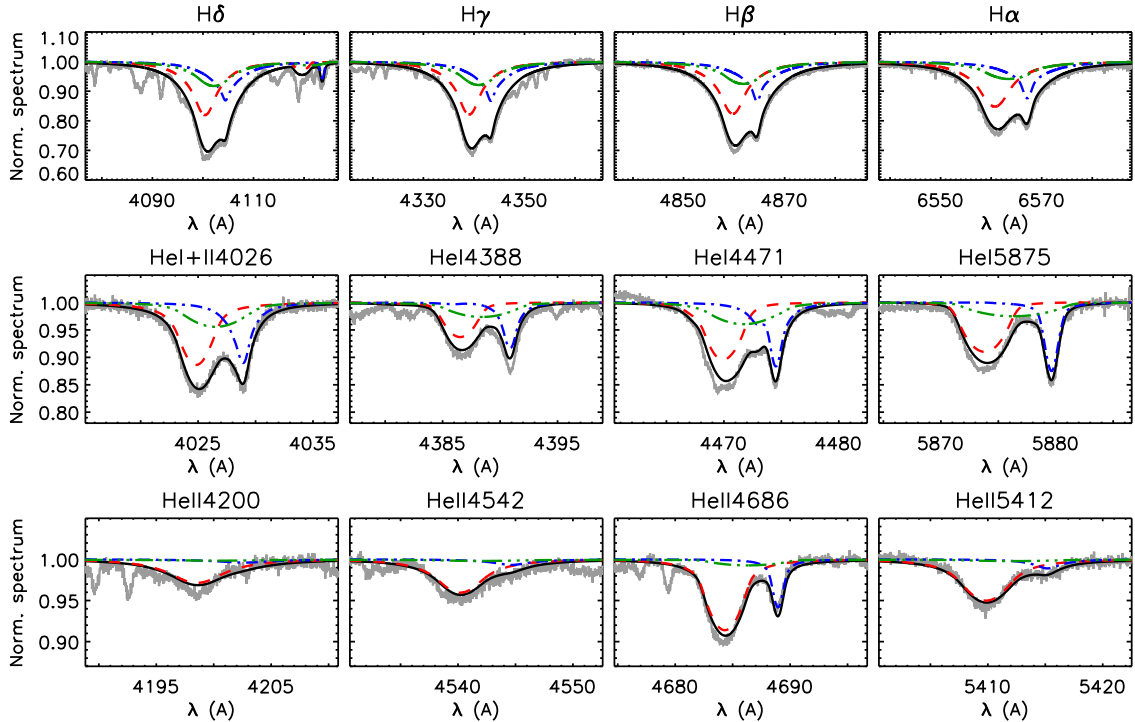}
      \caption{\footnotesize \fastwind\ analysis of the combined spectrum of $\sigma$~Ori~Aa,Ab,B at 
      periastron passage on 2013-10-31. Solid grey and black lines correspond to the observed spectrum
      and the best {\em combined} synthetic spectrum.
      The individual best-fit synthetic spectra for each component are overplotted with
      red dashed (Aa), blue dash-dotted (Ab), and green long-dash-dotted (B) lines.}
         \label{figure.FWfit}
   \end{figure*}

The best-fit solution was obtained by visual comparison of the original and combined synthetic spectra around the H and \ioni{He}{i-ii} lines, which are commonly assumed as a diagnosis for the determination of stellar parameters of O stars 
\citep[e.g.,][]{Her92, Her02, Rep04}. 
The fit is illustrated by Fig.~\ref{figure.FWfit}, which also shows the relative contribution of each component to the global spectrum. 
Although the contribution of the B component to the combined H and \ioni{He}{i} lines is very small, the presence of a third component is needed to fit the region between the lines of the Aa and Ab components.
On the other hand, the contribution of B to the \ioni{He}{ii} lines is negligible, as expected from its lower effective temperature and large \vsini.

Fig.~\ref{figure.FWfit} also illustrates the difficulty to determine precisely the gravity of the 
three components (from the wings of the hydrogen Balmer lines) and the effective temperature 
of the elusive B component (mainly constrained by the weak \ioni{He}{ii} lines). To improve this
situation, we followed an iterative strategy that accounted for the comparison of spectroscopic and 
evolutionary masses (Section~\ref{section.physical}).
The resulting spectroscopic parameters are summarized in Table~\ref{t4}. 
The derived \Teff\ and \grav\ are also complemented with the gravities corrected from centrifugal acceleration \citep[\grav$_{\rm c}$, computed following the procedure described in][]{Rep04}. 
%We limited the accuracy of the quoted \Teff\ and \grav\ values to 500\,K and 0.05, respectively (i.e., half of the step size of our grid).
%-------------------------------------------------------------
% Fig 6
%-------------------------------------------------------------
%-------------------------------------------------------------

\subsubsection{Physical parameters}
\label{section.physical}

% -------------------------------------------------------
% Table 4
% -------------------------------------------------------
%
\begin{table*}[t!]
\begin{center}
\caption{\footnotesize Stellar properties of $\sigma$~Ori~Aa, Ab, and B.}
\label{t4}
\begin{tabular}{l c c c l}    
\hline
\hline
\noalign{\smallskip}
Parameter         & $\sigma$~Ori~Aa                 & $\sigma$~Ori~Ab  & $\sigma$~Ori~B & Unit  \\
\hline
\noalign{\smallskip}
$V$             & \solu{4.61}{0.02} & \solu{5.20}{0.03}  & \solu{5.31}{0.04} & mag \\
$M_V$ & \solu{--3.49}{0.11}  & \solu{--2.90}{0.11} & \solu{--2.79}{0.12} & mag \\
\vsini  & \solu{135}{15} & \solu{35}{5} & \solu{250}{50} & \kms \\
\Teff          & \solu{35.0}{1.0} & \solu{31.0}{1.0} & \solu{29.0}{2.0}  & kK \\
\grav\              & \solu{4.20}{0.15} & \solu{4.20}{0.15} & \solu{4.15}{0.20}  & cm\,s$^{-2}$ \\
\grav$_{\rm c}$      & \solu{4.21}{0.15} & \solu{4.20}{0.15} & \solu{4.18}{0.20} & cm\,s$^{-2}$  \\
$R$   & \solu{5.6}{0.3}   & \solu{4.8}{0.3}   & \solu{5.0}{0.3}  & $R_{\odot}$\\
$\log{L/L_{\odot}}$ & \solu{4.62}{0.07} & \solu{4.27}{0.07} & \solu{4.20}{0.13} &  \\
$M_{\rm sp}$   & \solu{18}{7}     & \solu{13}{5}     & \solu{14}{7} & $M_{\odot}$ \\
\noalign{\smallskip}
$M_{\rm ev}^{(a)}$   & 20.0$^{+0.9}_{-1.0}$  & 14.6$^{+0.8}_{-0.6}$     & 13.6$^{+1.1}_{-0.9}$ &  $M_{\odot}$  \\
Age$^{(b)}$   & 0.3$^{+1.0}_{-0.3}$ & 0.9$^{+1.5}_{-0.9}$ & 1.9$^{+1.6}_{-1.9}$ & Ma \\    
$v_{\rm rot, ini}$  & 150$^{+60}_{-50}$ & 40$^{+43}_{-35}$ & 270$^{+86}_{-70}$ & \kms \\
\noalign{\smallskip}
%$M_{\rm dyn}~$sin$^3$~$i~^{(2)}$ & \solu{9.78}{0.07} & \solu{7.38}{0.05} & -- &  $M_{\odot}$ \\
$\log{Q({\rm H}^0)}$  & 47.92 & 47.03 & 46.34 & s$^{-1}$ \\
$\log{Q({\rm He}^0)}$  & 46.25 & 44.58 & 43.35 & s$^{-1}$  \\
\hline
\end{tabular}
\tablenotetext{(a)}{$M_{\rm ev}$ refers to present-day evolutionary masses; however, due to the youth and low mass-loss rates of the stars present-day and initial masses are equal.}
\tablenotetext{(b)}{The quoted uncertainties in the estimated ages may actually be considered as upper limits since {\sc bonnsai} computations performed for this study assume that there is no correlation between the uncertainties associated with \Teff, $\log{L}$, and \grav\ (see notes in Sect.~\ref{section.physical} and Fig.~\ref{figure.tracks}).}
\end{center}
\end{table*}

%-------------------------------------------------------------
% Fig 8
%-------------------------------------------------------------
   \begin{figure}[t!]
   \centering
   \includegraphics[width=0.49\textwidth]{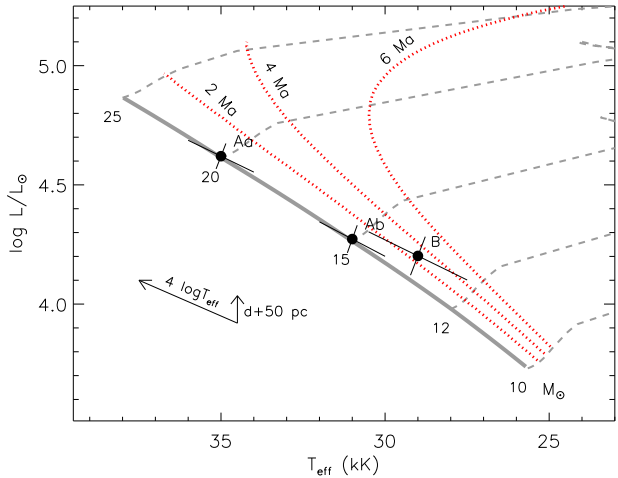}
      \caption{Hertzsprung-Russel diagram showing the three components of $\sigma$~Ori~Aa,Ab,B, and evolutionary tracks (grey {\em dashed}), ZAMS line (black solid), and 2, 4, and 6\,Ma isochrones for stellar masses in the range 10--25\,$M_\odot$ and solar metallicity from \citet{Bro11} models. 
Sizes of error bars of effective temperature and luminosity of $\sigma$~Ori~Aa, Ab, and B were computed as if both quantities were not correlated.
The diagonal and vertical arrows indicate the expected direction of correlated variability of $T_{\rm eff}$ and $L$, and of $L$ when heliocentric distance is increased by 50\,pc, respectively.}
         \label{figure.tracks}
   \end{figure}
%-------------------------------------------------------------

Once effective temperatures and gravities were determined, we used them, along with 
the spectral energy distributions of the associated \fastwind\ models and the absolute visual magnitudes computed
in Section~\ref{section.photometry}, to obtain the radii, luminosities and {\em spectroscopic} 
masses ($M_{\rm sp}$) of the three components. 
To this aim, we followed the same strategy as described in \citet{Her92}.
The resulting stellar parameters and associated uncertainties from {\sc fastwind} are given in Table \ref{t4}. 
The errors in $R$, log$L$, and $M_{\rm sp}$ take into account the uncertainties in distance, extinction, effective temperatures, gravities, and
individual extinction-corrected $V$-band absolute magnitudes (obtained in turn from the uncertainties in the combined apparent magnitude and 
magnitude differences between A and B, and Aa and Ab; see Section~\ref{section.photometry}).

We then used the Bayesian code {\sc bonnsai}\footnote{\tt http://www.astro.uni-bonn.de/stars/bonnsai} 
\citep{Sch14b} to infer the {\em evolutionary} masses ($M_{\rm ev}$) and ages of $\sigma$~Ori~Aa, Ab, and B. 
Once some priors are accounted for, {\sc bonnsai} matches all available observables simultaneously to stellar models and delivers probability distributions of the derive stellar parameters.
In our case, we matched the luminosities, surface gravities, effective temperatures, and projected rotational velocities of our three stars to the rotating Milky Way stellar models of \citet{Bro11}. 
We assumed a Salpeter initial mass function \citep{Sal55} as initial mass prior, uniform priors for age and initial rotational velocity, and that all rotation axes are randomly distributed in space. 
The stellar models reproduced correctly the observables.
The determined initial masses, ages and initial rotational velocities are summarized in Table~\ref{t4} including their $1\sigma$ uncertainties. 

For completeness, we also provide in Table~\ref{t4} the number of hydrogen (H$^0$) and neutral helium (He$^0$) ionizing photons emitted by $\sigma$~Ori~Aa, Ab, and B. 
These quantities can be of future interest for, e.g., the study of the ionization of the \object{Horsehead Nebula}, the associated photo-dissociation region, and the impact of massive stars on the formation and evolution of low-mass stars, brown dwarfs, stellar discs, Herbig-Haro objects, and possible planetary systems in the $\sigma$~Orionis star forming region. 
In total, the triple system emits 9.6$\times$10$^{47}$ H$^0$ and 1.8$\times$10$^{46}$\,s$^{-1}$ He$^0$ ionizing photons, enough for shaping virtually all remnant clouds in the Ori~OB1b association \citep{Ogu98}.
However, the ionization budget of the surrounding interstellar medium is mainly dominated by the hottest component only, $\sigma$~Ori~Aa, which contributes with 87\,\% (H$^0$) and 98\,\% (He$^0$) of the total numbers of ionizing photons.

%================================================================================
\section{Discussion}\label{section.discussion}
%================================================================================

\subsection{Summary of orbital properties of the $\sigma$~Ori~Aa,Ab,B system}\label{section.context}
\label{orbit}

As shown before, the angular separation between $\sigma$~Ori~Aa,Ab and B lies today at $\sim$0.25\,arcsec, which translates into a projected physical separation of about 100\,au \citep[][and references therein]{Cab14}.
This separation makes the ``pair'' to be resolvable only with the {\em Hubble Space Telescope} and, from the ground, with micrometers, speckle, adaptive optics, or lucky imagers at  $>$1\,m-class telescopes. 

The last published orbital solution, by \citet{Tur08}, quoted a period $P$ = 156.7$\pm$3.0\,a, an eccentricity $e$ = 0.0515$\pm$0.0080, and an inclination angle $i$ = 159.7$\pm$3.7\,deg (i.e., the orbit is almost circular and located in the plane of the sky). Our AstraLux astrometric measurements (Table~\ref{AstraLuxtable}), together with many other independent ones obtained after 2008, will certainly help to refine the published orbital solution.

Regarding the $\sigma$~Ori~Aa,Ab pair, over a century had to elapse since the first suspicion of spectroscopic binarity by \citet{Fro04}, through the first quantitative measures by \citet{Bol74} and \citet{Sti01}, to the confirmation by \citet{Sim11a}. 
With a projected physical separation between 0.8 and 6.4\,mas, the Aa,Ab system can be resolved in imaging only with interferometric observations. After publication of \cite{Sim11a}, and during the development of this paper, two different teams have been able to resolve the Aa,Ab pair with interferometric facilities \citep{Hum13, Sch13}.
%the Navy Precision Optical Interferometer \citep[NPOI][]{Hum13}, and the Michigan Infra-Red Combiner for the CHARA Interferometer \citep[MIRC][]{Sch13}. 
While they plan to publish
a joint analysis (D. Gies; priv. comm.), both groups have already indicated
%provided preliminary values for the
%orbital inclination angle, dynamical masses (see Sect.~\ref{discussion.mass}), and distance (see Sect.~\ref{sub.distance}) to the stellar system. In particular, \citet{Sch13} and \citet{Hum13} indicated 
57 and 56.3\,deg, respectively, as first estimates for the inclination orbit of the Aa,Ab system. Our study allows to provide
an independent {rough} estimation of the inclination of the Aa,Ab orbit based on the comparison of our own determinations of the projected {\em dynamical} masses and {\em evolutionary} masses (see Sections~\ref{section.analysis} and \ref{discussion.mass}). 
Considering the values of $M_{\rm dyn}$~sin$^3~i$ and $M_{\rm ev}$ for the Aa and Ab 
components indicated in Table~\ref{t4}, we obtained $i$ = 52.0$\pm$1.2 and 52.9$\pm$1.2\,deg, respectively. 
These values are in relatively good agreement with (but a bit smaller than) the preliminary estimations provided by the combination of spectroscopic and interferometric observations.

%The combination of outcomes from the two complementary works (spectroscopic and interferometric) will be crucial for the definite global characterization of the orbital properties of the $\sigma$~Ori~Aa,Ab pair, as well as for the precise determination of absolute dynamical masses of the Aa,Ab,B trio and distance to the $\sigma$~Orionis cluster. 
%While we wait for those results from interferometry, we can already provide an independent {rough} estimation of the inclination of the Aa,Ab orbit based on the comparison of our own determinations of the projected {\em dynamical} masses and {\em evolutionary} masses (see Sections~\ref{section.analysis} and \ref{discussion.mass}). 
%Considering the values of $M_{\rm dyn}$~sin$^3~i$ and $M_{\rm ev}$ for the Aa and Ab components provided in Table~\ref{t4}, we obtained $i$ = 52.0$\pm$1.2 and 52.9$\pm$1.2\,deg, respectively. These values are in relatively good agreement with the preliminary estimations provided by \citet{Sch13} and \citet{Hum13}: 57 and 56.3\,deg, respectively. 
All the quoted values of $i$ imply that the $\sigma$~Ori~Aa,Ab system is not eclipsing. 
We have confirmed this result using {\em Hipparcos} photometry: we did not find any indication of possible eclipses, which should occur at phases 0.02 and 0.95 and, by folding the light curve to periods close to 143 days. 
%%% <<< ZVEZDA

In the light of the  orbital properties and derived projected rotational velocities of $\sigma$~Ori~Aa,Ab,B
(Table~\ref{t4}), we conclude that the geometry of this young hierarchical triple system is rather complex. 
On one hand, the close pair revolves in a very eccentric orbit with an inclination $\sim$55\,deg, while the wide pair does it in an almost circular orbit and with an inclination of $\sim$160~deg. 
Therefore, the triple system is far from being co-planar. 
On the other hand, the projected rotational velocities of the Aa and Ab components are very different (135 and 35\,\kms, respectively).
In absence of direct information about the inclination angle of the rotational axes, we hypothesize that the spins of the two stars are not synchronized and/or the rotational axes are not aligned. 
A more thorough investigation of the global properties of $\sigma$~Ori~Aa,Ab,B in the context of the statistical properties, formation, and evolution of multiple stellar systems, as in \citet{Tok08}, is one possible direction of future work.

The good phase coverage of our spectroscopic observations and the great accuracy in the determination of the orbital period of the very eccentric binary $\sigma$~Ori~Aa,Ab (with an error of 7.2 minutes in 143.2 days) allowed us to identify precisely the future dates of periastron and apoastron passages until 2020. 
The passage dates shown in Table~\ref{table.dates} will certainly ease the investigation of, e.g., absidal motion, circularization, orbital period variation, and, especially, X-ray emission (see below).

\subsection{$\sigma$~Ori~Aa,Ab,B as an X-ray emitter}\label{discussion.xray}

For years, $\sigma$~Ori~Aa,Ab,B has been identified as the most luminous and softest X-ray source in the $\sigma$~Orionis cluster \citep{Ber94, San04, Ski08, Cab09, Cab10}.
From a comprehensive analysis of high-resolution {\em Chandra} X-ray imaging and spectroscopy of the triple system (none of the Aa,Ab,B components can be resolved by existing X-ray telescopes), \citet{Ski08} concluded that the measured X-rays properties of $\sigma$~Ori were in good agreement with model predictions for shocks distributed in a radiatively driven stellar wind. 
However, other possible emission mechanisms that could not be ruled out by that time were a magnetically-confined wind shock in the weak-field limit, and a sub-terminal speed-colliding wind system, under the hypothesis that $\sigma$~Ori~A had a putative spectroscopic companion, which had not been confirmed yet.
With our latest results, the estimated minimum and maximum separations between the Aa and Ab components is $\sim$66 and $\sim$528\,$R_\odot$ (about 10 and 100 stellar radii), respectively.
Although previous investigations of X-ray emission variability have resulted in negative evidences \citep{San04, Ski08, Cab10}, likely due to the relatively long period of 143.2\,d and high eccentricity, the colliding wind shock hypothesis, now the most probable scenario, can be tested during future periastron passages with either {\em Chandra} or {\em XMM-Newton}.

\subsection{Stellar masses}\label{discussion.mass}

We provided a rigorous determination of the {\em evolutionary} masses and ages of the three components with the {\sc bonnsai} tool in Section~\ref{section.fastwind}. Complementing these estimates, the Hertzsprung-Russell diagram in Fig.~\ref{figure.tracks} allows a quick visual estimation of their {\em evolutionary} masses and ages, their associated uncertainties, and how these quantities are modified when the estimated effective temperatures and assumed distance are modified.
We derived a total {\em evolutionary} mass $M_{\rm Aa+Ab+B}$ = \solu{48.2}{1.5}\,$M_{\odot}$, a value that is in good agreement with the total {\em dynamical} mass derived by means of the Kepler's third law assuming the Turner et~al. (2008) orbital parameters for the ``astrometric'' pair and a distance of 385~pc ($M_{\rm A+B}$=\,\solu{44}{7}\,$M_{\odot}$). 
We also found fair matches between the ratios of evolutionary masses (1.37$\pm$0.10) and of dynamical masses (1.325$\pm$0.006; Table~\ref{table.orbparameters}) of the Aa and Ab components. 
%Assisted by the iteration in surface gravities during the spectroscopic analysis (Section~\ref{section.fastwind}), there is as well a correspondence between the individual {\em spectroscopic} and {\em evolutionary} masses.
%Actually, this comparison indicates that  slightly larger values of \grav\ would be needed for $\sigma$~Ori~Aa and $\sigma$~Ori~Ab, but by an amount smaller that the half-size of our grid of synthetic spectra (0.05\,dex). 
Finally, our derived {\em evolutionary} masses for the inner pair are systematically larger than the values resulting from the analysis of the interferometric observations ($M_{\rm dyn, Aa}$\,=\,16.7 and $M_{\rm dyn, Ab}$\,=\,12.6~$M_{\odot}$; Hummel et ~al. 2013; G.~Schaefer, priv. comm.). This is a puzzling result since the effective temperatures that would lead to these {\em evolutionary} masses ($\sim$32 and 29~kK, respectively; see Fig.~\ref{figure.tracks}) are incompatible with the values indicated by the quantitative spectroscopic analysis  (see also notes in Sect.~\ref{discussion.age}). A possible solution to this relative mismatch between the {\em evolutionary} and {\em dynamical} masses could be related to the inclination angle resulting from the combined radial velocity and interferometric analysis. In particular, taking into account the preliminary dynamical masses of $\sigma$~Ori~Aa and $\sigma$~Ori~Ab computed by \cite{Hum13} and \cite{Sch13}, and the total 
Aa+Ab+B mass indicated above (44$\pm$7 $M_{\odot}$), the B component would be more massive than the Ab component, while
the later is more luminous. Using our evolutionary masses there is a better correspondence between mass
and luminosity regarding the three more massive components of $\sigma$~Ori. As indicated in Sect.~\ref{section.context}, an orbit with a slightly smaller inclination would solve this mistmatch between
dynamical masses and luminosity.

\subsection{Stellar ages}\label{discussion.age}

The age of the $\sigma$~Orionis cluster has been extensively investigated in the literature, mainly using low-mass stars \citep[e.g.,][]{Zap02, She04, She08, Cab07, May08}. 
The widely accepted cluster age interval is 2--4\,Ma, with extreme values reported at 1.5 and 8\,Ma. 
Our study provides an independent determination of the age of the three most massive stars in the cluster, namely $\sigma$~Ori~Aa, Ab, and B. 
The individual stellar ages indicated by {\sc bonnsai} are 0.3$^{+1.0}_{-0.3}$, 0.9$^{+1.5}_{-0.9}$, and 1.9$^{+1.6}_{-1.9}$ Ma, respectively. In view of these values, once could argue that the derived ages for the three components are
in agreement (within the uncertainties) and propose a mean age $\sim$1~Ma for the $\sigma$\,Ori\,Aa,Ab,B system.
This result confirms the youth of the $\sigma$\,Orionis cluster, but also points towards stellar ages of the most massive members (especially $\sigma$\,Ori\,Aa and $\sigma$\,Ori\,Ab) which are slightly younger than the commonly accepted cluster age. This later statement is reinforced when one takes into account that the
quoted uncertaities in the ages resulting from the {\sc bonnsai} analysis may be actually considered as
upper limits. The {\sc bonnsai} computations assume that the uncertainties in the three input parameters
(\Teff, $\log{L}$, and \grav) are independent; however, as illustrated in Fig.~\ref{figure.tracks} there
is a strong correlation (linked to basic priciples of stellar astrophysics) between variations in 
$\log{L}$ and \Teff. These variations follow an inclined line in the HR diagram almost parallel to the ZAMS and the 1--4~Ma isochrones in the 10--20~$M_{\odot}$ range. As a consequence the uncertainty in the derived age due to uncertainties related to the effective temperature determination are actually smaller than predicted by {\sc bonnsai}.

Being aware of the subtlety of the result about the difference between the age derived for $\sigma$~Ori~Aa, Ab ($\leq$1~Ma) and the commonly accepted cluster age (2--4~Ma) we also wanted to investigate further whether this non-coevality is a result of an underestimation of the uncertainties in the derived ages or could be used as an observational evidence of an actual physical process linked to the formation or evolution of these stars.

We first discarded it to be a consequence of an incorrrect determination of the stellar parameters (especially effective temperatures), cluster distance, or individual magnitudes. 
Any of these three possibilities can locate $\sigma$~Ori~Aa and $\sigma$~Ori~Ab in a reasonable way on the 2\,Ma isochrone. However, an effective temperature lower than 32\,kK for $\sigma$~Ori~Aa would deliver a bad fitting of the  \ioni{He}{i-ii} ionization equilibrium (the same argument applies, to a less extent, to $\sigma$~Ori~Ab), a further distance, of 430\,pc, would translate into an age of about 6\,Ma for $\sigma$~Ori~B, and the magnitude difference between $\sigma$~Ori~Aa,Ab and B necessary to put the three components simultaneously on the ZAMS is incompatible with observations ($\Delta V \sim$ 1.5\,mag).

The assumed metallicity and input physics considered in the stellar evolution models could also modify the derived ages. To check the effect of these parameters we compared the location of the tracks and isochrones computed by the Bonn \citep{Bro11} and Geneva \citep{Eks12} groups, which assumed a slightly different {\em solar} metallicity and core-overshooting parameter. The position of the ZAMS (for the same initial rotational velocity) in \cite{Eks12} models is exactly the same as for \cite{Bro11}; on the other hand, the isochrones of a given age in \cite{Eks12} computations are closer to the ZAMS compared to \cite{Bro11} models. As a consequence, Geneva models implied slightly older ages for $\sigma$~Ori~Aa and $\sigma$~Ori~Ab; however, these were still less than 1~Ma.

We also explored the possibility that $\sigma$~Ori~Aa and $\sigma$~Ori~Ab have suffered
from a rejuvenation process by mass accretion \citep[viz.][]{Bra95, van98, Dra07, Sch14a}.
As indicated in Sect.~\ref{section.context}, these two stars with masses $\sim$20 and 
$\sim$15~$M_{\odot}$ are orbiting in a high eccentric orbit with a period of 143.2 days. 
The closest separation between the stars in the orbit, at periastron, is $\sim$65\,$R_{\odot}$ (more than a factor ten in stellar radii). This wide separation, together with the small tabulated mass-loss rate \citep{Naj11}, makes the occurence of mass transfer episodes (and hence any associated 
rejuvenation phenomenon) in this system very unlikely. 

Are we then facing to an empirical evidence of a star formation process in which massive stars in a stellar cluster are formed slightly after their low-massive counterparts? The formation of massive stars
is still an open, highly debated question. \citet{Zin07} reviewed the three main competing concepts
of massive star formation, including (a) the monolitic collapse in isolated cores, (b) the competitive 
accretion in a protocluster environment, and (c) stellar collisions and mergers in very dense clusters. 
In absence of any other satisfactory solution at this point, we might assume the competitive accretion
scenario as a likely explanation for the slightly younger age derived for $\sigma$~Ori~Aa,Ab compared 
to cluster age as determined from low-mass stars. 
Although it must be considered for the moment as a highly speculative statement, its confirmation/refutation  (based on a much deeper study including all possible observational information regarding the high- and low-mass content of the $\sigma$~Orionis cluster and its viability in terms of spatial and temporal scales) deserves further investigation. In this context, we remark the recent study by \cite{Riv13} in which the authors claims that the scenario that better explain the distribution of pre-main sequence stars in three star formation regions in Orion (the Trapezium cluster, the Orion hot core, and the OMC1-S region) assumes high fragmentation in the parental core, accretion at subcore scales that forms a low-mass stellar cluster, and subsequent competitive accretion.

%================================================================================
\section{Summary}\label{section.summary}
%================================================================================

The almost extinction-free, young $\sigma$~Orionis cluster in the Ori~OB1b association is a 
cornerstone region for the study of stellar and substellar formation and the interplay 
between the strong (far-)UV radiation emitted by massive stars and the surrounding interstellar material.
At the very centre of the cluster, the massive $\sigma$~Ori~A,B system, commonly identified as
a close astrometric binary, has been recently confirmed to be a hierarchical triple system. The 
first two stars ($\sigma$~Ori Aa and Ab) are coupled in a very eccentric orbit with a period $\sim$143~d, 
and both together are orbiting with the third component ($\sigma$~Ori~B) in a much wider and 
longer ($P\,\sim$\,156.7 a) orbit almost circular.

Aiming at providing a complete characterization of the physical properties of the three components
and an improved set of orbital parameters for the highly eccentric {$\sigma$~Ori~Aa,Ab} pair
we compiled and analyzed a spectroscopic dataset comprising 94 high-resolution spectra ($\sim$33 of 
them were obtained near periastron passage of the Aa-Ab system). The complete sample covers a 
total time-span of almost 14 orbital periods of the $\sigma$~Ori~Aa,Ab system. 

The revised orbital analysis of the radial velocity curves of the $\sigma$~Ori~Aa,Ab pair led to 
an improved orbital solution compared to our previous study performed in \citet{Sim11a}. The great accuracy reached in the determination of the orbital period (7.2 minutes in 143.198 days) allowed us to provide precise future ephemerides for the system. This can be of particular interest for the investigation of variability of 
the strong X-ray emission detected for $\sigma$~Ori~A,B. In addition, the good phase coverage achieved 
by our observations settle a firm baseline for future investigations of apsidal motion effects, 
circularization and/or time variation of the orbital period in this young, very eccentric massive 
stellar system.

We performed a {\em combined} quantitative spectroscopic analysis of the $\sigma$~Ori~Aa,Ab,B system by means of the stellar atmosphere code {\sc fastwind}. We used own plus other available information about photometry and distance to the system to provide estimates for the radii, luminosities, and {\em spectroscopic} masses of the three components. We also infered {\em evolutionary} masses and stellar ages using the Bayesian code {\sc bonnsai}.

Despite the absence of clear spectroscopic features associated to the $\sigma$~Ori~B component in the combined spectrum, we provided indirect arguments indicating that the faintest star in the traditionally considered astrometric binary is an early-B type dwarf with a projected rotational velocity of at least 200~\kms. 
The \fastwind+\bonnsai\ analysis indicated that the $\sigma$~Ori~Aa,Ab pair contains the hottest \\
($T_{\rm eff, Aa}$\,=\solu{35.0}{1.0}~kK, $T_{\rm eff, Ab}$\,=\solu{31.0}{1.0}~kK) and most massive \\ 
($M_{\rm Aa}$\,=\,\solu{20.0}{1.0}~$M_{\odot}$, $M_{\rm Ab}$\,=\,\solu{14.6}{0.7}~$M_{\odot}$) components 
of the triple system, while $\sigma$~Ori~B is a bit cooler and less massive ($T_{\rm eff, B}$\,=\solu{29.0}{1.5}~kK, $M_{\rm B}$\,=\,\solu{13.6}{0.8}~$M_{\odot}$). The derived stellar age for $\sigma$~Ori~B
(1.9$^{+1.6}_{-1.9}$~Ma) is in relative good agreement with previous determinations of the age of 
the $\sigma$~Orionis cluster; however, the ages of the spectroscopic pair (0.3$^{+1.0}_{-0.3}$ and 0.9$^{+1.5}_{-0.9}$, respectively) are intriguingly younger than the commonly accepted cluster age (2--4~Ma).

The outcome of this study, once combined with on-going interferometric and past/future optical and X-ray observations, will be of key importance for a precise determination of the distance to the $\sigma$~Orionis cluster, the interpretation of the strong X-ray emission of the $\sigma$~Ori~Aa,Ab,B system, and the investigation of the formation and evolution of multiple massive stellar stystems and substellar objects.

%================================================================================

\acknowledgments

This research made use of the SIMBAD, operated at Centre de Donn\'ees astronomiques de Strasbourg, France, and NASA's Astrophysics Data System.
Financial support was provided by the Spanish Ministerio de Ciencia e Innovaci\'on and Ministerio de Econom\'{\i}a y Competitividad under grants 
AYA2010-17631, % Jesus 1
AYA2010-15081, % Jesus 2
AYA2010-21697-C05-04, % Sergio 1
AYA2011-30147-C03-03,	% (CARMENES-CAB) 
AYA2012-39364-C02-01/02, % Sergio 2
and Severo Ochoa SEV-2011-0187, and 
by the Canary Islands Government under grant PID2010119.
RHB acknowledges support from FONDECYT Regular project 1140076.
JSB acknowledges support by the “JAE-PreDoc” program of the Spanish Consejo Superior de 
Investigaciones Científicas (CSIC).
\'A.S. acknowledges support by the J\'anos Bolyai Research Scholarship of the Hungarian Academy of Sciences. 
We acknowledge F. Najarro, J. Puls, A. Herrero, J. Casares, and H. Blau for interesting discussions during the developement of this paper and their useful comments to one of the latest drafts before submission.
We thank G. Schaefer for providing us some information about the results from the analysis of the interferometric observations prior to publication. Last, to the anonymous referee for a very professional report that help us to improve the first version of the manuscript.\\
%{\bf Lo mismo para FIES, HERMES, CAFE, FEROS o quitamos HRS (sugiero lo segundo).}
 Based on observations made with the Nordic Optical Telescope, operated by the Nordic Optical Telescope Scientific Association at the Observatorio del Roque de los Muchachos, La Palma, Spain, of the Instituto de Astrofisica de Canarias.
 Based on observations made with the Mercator Telescope, operated on the island of La Palma by the Flemish Community, at the Spanish Observatorio del Roque de los Muchachos of the Instituto de Astrofísica de Canarias. 
 Based on observations obtained with the HERMES spectrograph, which is supported by the Fund for Scientific Research of Flanders (FWO), Belgium, the Research Council of K.U.Leuven, Belgium, the Fonds National de la Recherche Scientifique (F.R.S.-FNRS), Belgium, the Royal Observatory of Belgium, the Observatoire de Genève, Switzerland and the Thüringer Landessternwarte Tautenburg, Germany. 
The Hobby-Eberly Telescope (HET) is a joint project of the University of Texas at Austin, the Pennsylvania State University, Stanford University, Ludwig-Maximilians-Universit\"at M\"unchen, and Georg-August-Universit\"at G\"ottingen. The HET is named in honor of its principal benefactors, William P. Hobby and Robert E. Eberly.

%% To help institutions obtain information on the effectiveness of their
%% telescopes, the AAS Journals has created a group of keywords for telescope
%% facilities. A common set of keywords will make these types of searches
%% significantly easier and more accurate. In addition, they will also be
%% useful in linking papers together which utilize the same telescopes
%% within the framework of the National Virtual Observatory.
%% See the AASTeX Web site at http://aastex.aas.org/
%% for information on obtaining the facility keywords.

%% After the acknowledgments section, use the following syntax and the
%% \facility{} macro to list the keywords of facilities used in the research
%% for the paper.  Each keyword will be checked against the master list during
%% copy editing.  Individual instruments or configurations can be provided 
%% in parentheses, after the keyword, but they will not be verified.

{\it Facilities:} \facility{NOT}, \facility{Mercator1.2m}, \facility{CAO:2.2m}, \facility{HET}, \facility{Max Planck:2.2m}.

\clearpage

{\small
\begin{longtable}{ccc cc cc l}
\label{table.rvs}\\
\caption[]{Radial velocity measurements of $\sigma$~Ori~Aa and~Ab.}\\
   \hline
   \hline
   \noalign{\smallskip}
Date		& HJD 		 & $\phi$ & $V_r$ (Aa)	& $O-C$ (Aa) & $V_r$ (Ab)	& $O-C$ (Ab) & Instrument \\
(yyyy-mm-dd)        	& (--2450000)	 &	  & [\kms]  	& [\kms]	& [\kms] & [\kms] & \\
\noalign{\smallskip}
    \hline
    \noalign{\smallskip}		
 \endfirsthead
  \caption[]{Radial velocity measurements of $\sigma$~Ori~Aa and~Ab (cont.).}\\ 
  \hline
  \hline
  \noalign{\smallskip}		
Date		& HJD 		 & $\phi$ & $V_r$ (Aa)	& $O-C$ (Aa) & $V_r$ (Ab)	& $O-C$ (Ab) & Instrument \\
(yyyy-mm-dd)        	& (--2450000)	 &	  & [\kms]  	& [\kms]	& [\kms] & [\kms] & \\
 \noalign{\smallskip}
  \hline
  \noalign{\smallskip}
  \endhead
  \noalign{\smallskip}
  \hline
  \endfoot
  2008-11-05 &  4776.735 & 0.284 &    50.1 $\pm$    3.5 &    --0.2 &     6.0 $\pm$    1.9 &     0.3 &    FIES \\
  2008-11-05 &  4776.740 & 0.284 &    50.2 $\pm$    3.6 &    --0.1 &     7.0 $\pm$    1.9 &     1.4 &    FIES \\
  2008-11-06 &  4777.697 & 0.291 &    50.4 $\pm$    3.6 &     0.1 &     6.2 $\pm$    2.0 &     0.5 &    FIES \\
  2008-11-06 &  4777.699 & 0.291 &    50.3 $\pm$    3.6 &     0.0 &     6.3 $\pm$    2.0 &     0.6 &    FIES \\
  2008-11-06 &  4777.700 & 0.291 &    50.4 $\pm$    3.5 &     0.1 &     5.9 $\pm$    1.9 &     0.2 &    FIES \\
  2008-11-07 &  4778.718 & 0.298 &    49.1 $\pm$    3.5 &    --1.1 &     8.0 $\pm$    1.9 &     2.2 &    FIES \\
  2008-11-07 &  4778.719 & 0.298 &    49.3 $\pm$    3.5 &    --0.9 &     7.6 $\pm$    2.0 &     1.8 &    FIES \\
  2008-11-08 &  4779.704 & 0.305 &    49.1 $\pm$    3.5 &    --1.0 &     6.3 $\pm$    2.0 &     0.4 &    FIES \\
  2008-11-08 &  4779.705 & 0.305 &    46.9 $\pm$    3.6 &    --3.2 &     7.6 $\pm$    1.9 &     1.7 &    FIES \\
  2009-05-02 &  4953.965 & 0.522 &    43.9 $\pm$    3.4 &    --1.5 &    12.0 $\pm$    1.9 &    --0.2 &   FEROS \\
  2009-05-03 &  4954.963 & 0.529 &    43.0 $\pm$    3.4 &    --2.2 &    12.8 $\pm$    2.0 &     0.3 &   FEROS \\
  2009-11-09 &  5145.659 & 0.860 &    18.2 $\pm$    3.5 &    --0.5 &    48.0 $\pm$    1.9 &     0.5 &    FIES \\
  2009-11-11 &  5147.650 & 0.874 &    14.4 $\pm$    3.4 &    --1.2 &    52.6 $\pm$    1.9 &     0.9 &    FIES \\
  2010-09-07 &  5447.730 & 0.970 &   --45.8 $\pm$    1.3 &     0.5 &   135.0 $\pm$    1.0 &     1.3 &    FIES \\
  2010-09-09 &  5449.747 & 0.984 &   --74.2 $\pm$    1.5 &     1.2 &   173.7 $\pm$    1.2 &     1.5 &    FIES \\
  2010-10-21 &  5490.729 & 0.270 &    46.1 $\pm$    3.5 &    --4.3 &     8.6 $\pm$    1.9 &     3.0 &  HERMES \\
  2010-10-23 &  5493.734 & 0.291 &    50.7 $\pm$    3.5 &     0.4 &     6.2 $\pm$    2.0 &     0.5 &    FIES \\
  2010-10-23 &  5493.736 & 0.291 &    51.5 $\pm$    3.4 &     1.2 &     5.8 $\pm$    1.9 &     0.1 &    FIES \\
  2011-01-11 &  5573.509 & 0.848 &    21.5 $\pm$    3.5 &     0.4 &    43.8 $\pm$    2.0 &    --0.6 &    FIES \\
  2011-01-15 &  5577.505 & 0.876 &    13.8 $\pm$    3.4 &    --1.3 &    52.6 $\pm$    2.0 &     0.3 &    FIES \\
  2011-01-15 &  5577.509 & 0.876 &    15.1 $\pm$    3.4 &    --0.0 &    52.6 $\pm$    1.9 &     0.3 &    FIES \\
  2011-01-15 &  5577.512 & 0.876 &    14.5 $\pm$    3.5 &    --0.6 &    52.7 $\pm$    2.0 &     0.4 &    FIES \\
  2011-02-11 &  5604.367 & 0.064 &    34.4 $\pm$    3.5 &     1.2 &    28.3 $\pm$    2.0 &     0.0 &    FIES \\
  2011-02-13 &  5606.027 & 0.075 &    36.1 $\pm$    3.4 &    --1.5 &    21.9 $\pm$    2.0 &    --0.6 &   FEROS \\
  2011-02-20 &  5613.359 & 0.127 &    50.3 $\pm$    3.3 &     3.7 &     9.3 $\pm$    2.0 &    --1.3 &    FIES \\
  2011-03-22 &  5643.024 & 0.334 &    50.0 $\pm$    3.4 &     0.2 &     5.8 $\pm$    1.9 &    --0.5 &   FEROS \\
  2011-03-27 &  5648.379 & 0.371 &    52.5 $\pm$    3.4 &     3.3 &     5.1 $\pm$    1.9 &    --2.0 &    FIES \\
  2011-04-08 &  5660.353 & 0.455 &    47.1 $\pm$    3.4 &    --0.3 &     7.8 $\pm$    2.0 &    --1.8 &    FIES \\
  2011-09-07 &  5812.724 & 0.519 &    43.2 $\pm$    3.3 &    --2.3 &    15.2 $\pm$    2.0 &     3.2 &    FIES \\
  2011-09-08 &  5813.743 & 0.526 &    43.7 $\pm$    3.3 &    --1.5 &    16.4 $\pm$    2.0 &     4.1 &    FIES \\
  2011-09-09 &  5814.717 & 0.533 &    43.8 $\pm$    3.5 &    --1.2 &    14.4 $\pm$    2.0 &     1.8 &    FIES \\
  2011-09-10 &  5815.726 & 0.540 &    43.2 $\pm$    3.5 &    --1.6 &    13.6 $\pm$    2.0 &     0.6 &    FIES \\
  2011-09-11 &  5816.748 & 0.547 &    42.4 $\pm$    3.5 &    --2.1 &    15.1 $\pm$    2.0 &     1.8 &    FIES \\
  2011-09-12 &  5817.758 & 0.554 &    41.4 $\pm$    3.5 &    --2.9 &    15.8 $\pm$    1.9 &     2.2 &    FIES \\
  2011-10-04 &  5838.961 & 0.702 &    36.7 $\pm$    3.4 &    --0.2 &    27.0 $\pm$    1.9 &     3.6 &     HRS \\
  2011-11-09 &  5874.575 & 0.951 &   --17.6 $\pm$    2.6 &     3.7 &   101.1 $\pm$    1.4 &     0.7 &  HERMES \\
  2011-11-16 &  5881.847 & 0.001 &   --82.8 $\pm$    1.2 &     2.9 &   184.7 $\pm$    1.1 &    --1.1 &     HRS \\
  2011-11-18 &  5883.824 & 0.015 &   --29.5 $\pm$    2.3 &     5.3 &   121.7 $\pm$    1.2 &     3.2 &     HRS \\
  2011-11-22 &  5887.815 & 0.043 &    19.4 $\pm$    3.4 &    --0.3 &    46.7 $\pm$    2.0 &     0.5 &     HRS \\
  2011-11-24 &  5889.821 & 0.057 &    29.8 $\pm$    3.6 &    --0.2 &    33.4 $\pm$    1.9 &     0.8 &     HRS \\
  2012-01-03 &  5929.710 & 0.336 &    52.0 $\pm$    3.6 &     2.2 &     5.6 $\pm$    1.9 &    --0.8 &     HRS \\
  2012-04-05 &  6023.388 & 0.990 &   --86.4 $\pm$    1.8 &     1.4 &   185.6 $\pm$    1.0 &    --3.0 &  HERMES \\
  2012-04-05 &  6023.393 & 0.990 &   --87.4 $\pm$    2.0 &     0.5 &   185.9 $\pm$    1.1 &    --2.8 &  HERMES \\
  2012-04-06 &  6024.329 & 0.996 &   --94.3 $\pm$    1.8 &    --1.1 &   193.4 $\pm$    1.4 &   --2.3 &  HERMES \\
  2012-04-06 &  6024.333 & 0.997 &   --93.1 $\pm$    1.2 &     0.1 &   192.9 $\pm$    1.1 &    --2.8 &  HERMES \\
  2012-04-06 &  6024.344 & 0.997 &   --95.6 $\pm$    1.4 &    --2.5 &   196.4 $\pm$    1.3 &     0.7 &  CAF\'E \\
  2012-04-07 &  6025.314 & 0.003 &   --80.7 $\pm$    1.3 &    --0.5 &   178.3 $\pm$    1.2 &   --0.2 &  CAF\'E \\
  2012-04-08 &  6026.321 & 0.010 &   --52.2 $\pm$    1.6 &     1.1 &   142.9 $\pm$    0.9 &    --0.1 &  CAF\'E \\
  2012-04-08 &  6026.326 & 0.010 &   --53.6 $\pm$    1.8 &    --0.4 &   145.0 $\pm$    0.9 &     2.2 &  HERMES \\
  2012-10-26 &  6226.679 & 0.410 &    46.5 $\pm$    3.3 &    --2.0 &     9.5 $\pm$    1.9 &     1.4 &  HERMES \\
  2012-10-27 &  6227.659 & 0.416 &    44.6 $\pm$    3.5 &    --3.7 &    12.2 $\pm$    1.9 &     3.9 &  HERMES \\
  2012-10-29 &  6229.584 & 0.430 &    44.7 $\pm$    3.5 &    --3.3 &    11.5 $\pm$    1.9 &     2.8 &  HERMES \\
  2012-12-23 &  6285.551 & 0.821 &    25.2 $\pm$    3.5 &    --0.3 &    38.7 $\pm$    2.0 &     0.2 &    FIES \\
  2012-12-24 &  6286.599 & 0.828 &    26.1 $\pm$    3.5 &     1.7 &    35.4 $\pm$    2.0 &    --4.5 &    FIES \\
  2012-12-25 &  6287.467 & 0.834 &    23.0 $\pm$    3.5 &    --0.5 &    37.1 $\pm$    2.0 &    --4.1 &    FIES \\
  2013-01-28 &  6321.409 & 0.071 &    36.1 $\pm$    3.5 &    --0.1 &    24.4 $\pm$    2.0 &     0.0 &    FIES \\
  2013-01-29 &  6322.470 & 0.079 &    40.2 $\pm$    3.5 &     1.6 &    21.7 $\pm$    2.0 &     0.5 &    FIES \\
  2013-01-30 &  6323.396 & 0.085 &    38.8 $\pm$    3.4 &    --1.5 &    19.6 $\pm$    2.0 &     0.6 &    FIES \\
  2013-02-05 &  6329.389 & 0.127 &    47.1 $\pm$    3.5 &     0.5 &    13.6 $\pm$    1.9 &     3.0 &    FIES \\
  2013-02-05 &  6329.391 & 0.127 &    48.2 $\pm$    3.5 &     1.6 &    11.4 $\pm$    1.9 &     0.8 &    FIES \\
  2013-02-15 &  6339.330 & 0.196 &    50.3 $\pm$    3.4 &     0.4 &     3.5 $\pm$    1.9 &    --2.7 &    FIES \\
  2013-03-09 &  6361.470 & 0.351 &    49.2 $\pm$    3.5 &    --0.3 &     5.1 $\pm$    1.9 &    --1.5 &  HERMES \\
  2013-03-09 &  6361.473 & 0.351 &    49.2 $\pm$    3.5 &    --0.3 &     5.0 $\pm$    1.9 &    --1.6 &  HERMES \\
  2013-03-10 &  6362.459 & 0.358 &    49.9 $\pm$    3.4 &     0.5 &     4.5 $\pm$    1.9 &    --2.3 &  HERMES \\
  2013-10-25 &  6590.638 & 0.951 &   --18.9 $\pm$    2.6 &     2.9 &   103.3 $\pm$    1.3 &     2.2 &  HERMES \\
  2013-10-26 &  6591.627 & 0.958 &   --28.7 $\pm$    2.3 &     0.6 &   114.2 $\pm$    1.3 &     3.1 &  HERMES \\
  2013-10-27 &  6592.627 & 0.965 &   --36.2 $\pm$    2.2 &     2.5 &   124.8 $\pm$    1.2 &     1.3 &  HERMES \\
  2013-10-28 &  6593.585 & 0.972 &   --52.2 $\pm$    1.3 &    --2.5 &   138.0 $\pm$    0.9 &   --0.2 &  HERMES \\
  2013-10-29 &  6594.697 & 0.980 &   --65.0 $\pm$    1.4 &     0.5 &   160.4 $\pm$    1.0 &     1.4 &  HERMES \\
  2013-10-30 &  6595.768 & 0.987 &   --83.3 $\pm$    1.4 &    --1.1 &   180.6 $\pm$    1.4 &   --0.6 &  HERMES \\
  2013-10-31 &  6596.596 & 0.993 &   --93.4 $\pm$    1.3 &    --1.5 &   197.1 $\pm$    1.3 &     3.0 &  HERMES \\
  2013-10-31 &  6596.777 & 0.994 &   --93.5 $\pm$    1.4 &    --0.6 &   197.9 $\pm$    1.2 &     2.5 &  HERMES \\
  2013-11-01 &  6597.783 & 0.001 &   --88.9 $\pm$    1.3 &    --2.3 &   188.2 $\pm$    1.1 &     1.2 &  HERMES \\
  2013-11-02 &  6598.787 & 0.008 &   --60.9 $\pm$    1.6 &     1.5 &   157.0 $\pm$    0.9 &     2.0 &  HERMES \\
  2013-11-04 &  6600.789 & 0.022 &   --13.7 $\pm$    2.8 &     0.5 &    92.2 $\pm$    1.6 &     1.1 &  HERMES \\
  2013-11-05 &  6601.787 & 0.029 &     0.3 $\pm$    3.3 &    --0.7 &    71.1 $\pm$    1.9 &     0.1 &  HERMES \\
  2013-11-06 &  6602.797 & 0.036 &    13.6 $\pm$    3.2 &     1.8 &    52.9 $\pm$    1.9 &    --3.7 &  HERMES \\
  2013-11-07 &  6603.788 & 0.043 &    18.9 $\pm$    3.4 &    --0.7 &    45.5 $\pm$    1.9 &    --0.9 &  HERMES \\
  2013-11-08 &  6604.798 & 0.050 &    23.1 $\pm$    3.4 &    --2.3 &    39.8 $\pm$    1.9 &     1.2 &  HERMES \\
  2013-11-09 &  6605.800 & 0.057 &    27.6 $\pm$    3.4 &    --2.3 &    33.0 $\pm$    1.9 &     0.3 &  HERMES \\
  2013-11-10 &  6606.798 & 0.064 &    31.0 $\pm$    3.5 &    --2.4 &    28.5 $\pm$    1.9 &     0.4 &  HERMES \\
  2013-11-11 &  6607.777 & 0.071 &    32.8 $\pm$    3.3 &    --3.3 &    25.5 $\pm$    1.9 &     1.0 &  HERMES \\
  2013-11-29 &  6626.607 & 0.202 &    51.2 $\pm$    3.5 &     1.2 &     5.5 $\pm$    1.9 &    --0.5 &  CAF\'E \\
  2014-01-11 &  6669.455 & 0.502 &    45.5 $\pm$    3.5 &    --0.5 &    11.9 $\pm$    1.9 &     0.6 &  CAF\'E \\
  2014-03-10 &  6727.368 & 0.906 &     2.4 $\pm$    3.4 &    --3.5 &    66.2 $\pm$    1.8 &     1.7 &  CAF\'E \\
  2014-03-11 &  6728.376 & 0.913 &    --0.8 $\pm$    3.4 &    --3.8 &    68.6 $\pm$    1.9 &     0.3 &  CAF\'E \\
  2014-03-20 &  6737.401 & 0.976 &   --60.1 $\pm$    1.8 &    --2.0 &   150.0 $\pm$    0.9 &     0.7 &  CAF\'E \\
  2014-03-21 &  6738.308 & 0.982 &   --72.7 $\pm$    1.5 &    --0.8 &   167.1 $\pm$    1.1 &   --0.5 &  CAF\'E \\
  2014-03-22 &  6739.285 & 0.989 &   --87.0 $\pm$    1.5 &    --0.3 &   185.7 $\pm$    1.3 &   --1.4 &  CAF\'E \\
  2014-03-23 &  6740.290 & 0.996 &   --96.3 $\pm$    1.1 &    --3.0 &   194.3 $\pm$    1.4 &   --1.5 &  CAF\'E \\
\end{longtable}
}

\clearpage

\clearpage
% -------------------------------------------------------
% Table 5
% -------------------------------------------------------
\begin{table*}
\centering
\caption{Next epochs of quadratures of the $\sigma$~Ori~Aa,Ab pair.}
\label{table.dates}
\begin{tabular}{ccccc}
\hline
\hline
\noalign{\smallskip}
\multicolumn{2}{c}{Periastron}     &   & \multicolumn{2}{c}{Apastron} \\
\cline{1-2}  \cline{4-5}
\noalign{\smallskip}
Date        		& HJD  		& & Date			& HJD \\
(yyyy-mm-dd)   & (--2\,450\,000) 	& & (yyyy-mm-dd) 	& (--2\,450\,000) \\
\noalign{\smallskip}
\hline
\noalign{\smallskip}
 2015-01-04 &\solu{7027.22}{0.04}       & & 2015-03-17 &\solu{7098.82}{0.04}  \\
 2015-05-27$^*$ &\solu{7170.42}{0.04} & &   2015-08-07$^*$ &\solu{7242.01}{0.04}  \\
 2015-10-18 &\solu{7313.61}{0.05}       & & 2015-12-28 &\solu{7385.21}{0.05}  \\
 2016-03-09 &\solu{7456.81}{0.05}       & & 2016-05-19$^*$ &\solu{7528.41}{0.05}  \\
 2016-07-30$^*$ &\solu{7600.01}{0.06} & & 2016-10-10 &\solu{7671.61}{0.06}  \\
 2016-12-20 &\solu{7743.21}{0.06}       & & 2017-03-02 &\solu{7814.81}{0.06}  \\
 2017-05-12$^*$ &\solu{7886.40}{0.07}  & &  2017-07-23$^*$ &\solu{7958.00}{0.07}  \\
 2017-10-03 &\solu{8029.60}{0.07}       & & 2017-12-13 &\solu{8101.20}{0.07}  \\
 2018-02-23 &\solu{8172.80}{0.08}       & & 2018-05-05$^*$ &\solu{8244.40}{0.08}  \\
 2018-07-16$^*$ &\solu{8316.00}{0.08}  & & 2018-09-26 &\solu{8387.600}{0.08}  \\
 2018-12-06 &\solu{8459.20}{0.09}       & & 2019-02-16 &\solu{8530.80}{0.09}  \\
 2019-04-28$^*$ &\solu{8602.40}{0.09}  & &  2019-07-09$^*$ &\solu{8673.99}{0.09}  \\
 2019-09-19 &\solu{8745.59}{0.10}       & & 2019-11-29 &\solu{8817.19}{0.10}  \\
 2020-02-09 &\solu{8888.79}{0.10}       & & 2020-04-20$^*$ &\solu{8960.39}{0.10}  \\
 2020-07-01$^*$ &\solu{9031.99}{0.11}  & & 2020-09-11 &\solu{9103.59}{0.11}  \\
 2020-11-21 &\solu{9175.19}{0.11}       & & 2012-02-01 &\solu{9246.79}{0.11}  \\
\noalign{\smallskip}
\hline
\tablecomments{Dates marked with an asterisc denote that $\sigma$~Ori is not observable due its the proximity to the Sun.}
\end{tabular}
\end{table*}


\begin{thebibliography}{}
\bibitem[Abergel et~al.(2003)]{Abe03} Abergel, A., Teyssier, D., Bernard, J.~P., et~al.\ 2003, \aap, 410, 577 
\bibitem[Aceituno et~al.(2013)]{Ace13} Aceituno, J., S{\'a}nchez, S.~F., Grupp, F., et~al.\ 2013, \aap, 552, A31 
\bibitem[Barb{\'a} et~al.(2010)]{Bar10} Barb{\'a}, R.~H., Gamen, R., Arias, J.~I., et~al.\ 2010, Revista Mexicana de Astronom\'ia y Astrof\'isica Conference Series, 38, 30 
\bibitem[B\'ejar et~al.(1999)]{Bej99} B\'ejar, V. J. S., Zapatero Osorio, M. R., Rebolo, R. et~al. 1999, ApJ, 521, 671
\bibitem[Berghoefer \& Schmitt(1994)]{Ber94} Berghoefer, T.~W., \& Schmitt, J.~H.~M.~M.\ 1994, \aap, 290, 435 
\bibitem[Bolton(1974)]{Bol74} Bolton, C.~T.\ 1974, \apjl, 192, L7 
\bibitem[Bouy et~al.(2009)]{Bou09} Bouy, H., Hu{\'e}lamo, N., Mart{\'{\i}}n, E.~L., et~al.\ 2009, \aap, 493, 931 
\bibitem[Bowler et~al.(2009)]{Bow09} Bowler, B.~P., Waller, W.~H., Megeath, S.~T., Patten, B.~M., \& Tamura, M.\ 2009, \aj, 137, 3685 
\bibitem[Braun \& Langer(1995)]{Bra95} Braun, H., \& Langer, N.\ 1995, \aap, 297, 483 
\bibitem[Brott et~al.(2011)]{Bro11} Brott, I., de Mink, S.~E., Cantiello, M., et~al.\ 2011, \aap, 530, A115 
\bibitem[Brown et~al.(1994)]{Bro94} Brown, A.~G.~A., de Geus, E.~J., \& de Zeeuw, P.~T.\ 1994, \aap, 289, 101 
\bibitem[Burnham(1892)]{Bur1892} Burnham, S. W. 1892, AN, 130, 257
\bibitem[Caballero(2005)]{Cab05} Caballero, J.~A. 2005, AN, 326, 1007 % ok
\bibitem[Caballero(2007)]{Cab07} Caballero, J.~A.\ 2007, \aap, 466, 917 % ok
\bibitem[Caballero(2008a)]{Cab08a} Caballero, J.~A. 2008a, MNRAS, 383, 750 % ok Dynamical parallax
\bibitem[Caballero(2008b)]{Cab08b} Caballero, J.~A. 2008b, A\&A, 487, 667 % ok Mayrit
\bibitem[Caballero(2013)]{Cab13} Caballero, J.~A. 2013, The Star Formation Newsletter, 243, 6  % ok
\bibitem[Caballero(2014)]{Cab14} Caballero, J.~A.\ 2014, Obs, 134, 273  %ok
\bibitem[Caballero \& Dinis(2008)]{CabDin08} Caballero, J.~A., \& Dinis, L.\ 2008, AN, 329, 801 %ok 
\bibitem[Caballero et~al.(2009)]{Cab09} Caballero, J.~A., L{\'o}pez-Santiago, J., de Castro, E., \& Cornide, M.\ 2009, \aj, 137, 5012 
\bibitem[Caballero et~al.(2010)]{Cab10} Caballero, J.~A., Albacete-Colombo, J.~F., \& L{\'o}pez-Santiago, J.\ 2010, \aap, 521, A45 
\bibitem[Compi{\`e}gne et~al.(2007)]{Com07} Compi{\`e}gne, M., Abergel, A., Verstraete, L., et~al.\ 2007, \aap, 471, 205 
\bibitem[de Zeeuw et~al.(1999)]{deZ99} de Zeeuw, P.~T., Hoogerwerf, R., de Bruijne, J.~H.~J., Brown, A.~G.~A., \& Blaauw, A.\ 1999, \aj, 117, 354 
\bibitem[Dray \& Tout(2007)]{Dra07} Dray, L.~M., \& Tout, C.~A.\ 2007, \mnras, 376, 61 
\bibitem[Ducati et al.(2001)]{Duc01} Ducati, J.~R., Bevilacqua, C.~M., Rembold, S.~B., \& Ribeiro, D.\ 2001, \apj, 558, 309 
\bibitem[Edwards(1976)]{Edw76} Edwards, T.~W.\ 1976, \aj, 81, 245 
\bibitem[Ekstr{\"o}m et al.(2012)]{Eks12} Ekstr{\"o}m, S., Georgy, C., Eggenberger, P., et al.\ 2012, \aap, 537, A146 
\bibitem[Frost \& Adams(1904)]{Fro04} Frost, E.~B., \& Adams, W.~S.\ 1904, \apj, 19, 151 
\bibitem[Garrison(1967)]{Gar67} Garrison, R. F. 1967, PASP, 79, 433
\bibitem[Goicoechea et~al.(2006)]{Goi06} Goicoechea, J.~R., Pety, J., Gerin, M., et~al.\ 2006, \aap, 456, 565 
\bibitem[Goicoechea et~al.(2009)]{Goi09} Goicoechea, J.~R., Pety, J., Gerin, M., Hily-Blant, P., \& Le Bourlot, J.\ 2009, \aap, 498, 771 
\bibitem[Gonz{\'a}lez Hern{\'a}ndez et~al.(2008)]{Gon08} Gonz{\'a}lez Hern{\'a}ndez, J.~I., Caballero, J.~A., Rebolo, R., et~al.\ 2008, \aap, 490, 1135 
\bibitem[Groote \& Hunger(1997)]{Gro97} Groote, D., \& Hunger, K.\ 1997, \aap, 319, 250 
\bibitem[Habart et~al.(2005)]{Hab05} Habart, E., Abergel, A., Walmsley, C.~M., Teyssier, D., \& Pety, J.\ 2005, \aap, 437, 177 
\bibitem[Hartkopf et~al.(1996)]{Har96} Hartkopf, W.~I., Mason, B.~D., \& McAlister, H.~A.\ 1996, \aj, 111, 370 
\bibitem[Hern{\'a}ndez et~al.(2005)]{Her05} Hern{\'a}ndez, J., Calvet, N., Hartmann, L., et~al.\ 2005, \aj, 129, 856 
\bibitem[Herrero et~al.(1992)]{Her92} Herrero, A., Kudritzki, R.~P., Vilchez, J.~M., et~al.\ 1992, \aap, 261, 209 
\bibitem[Herrero et~al.(2002)]{Her02} Herrero, A., Puls, J., \& Najarro, F.\ 2002, \aap, 396, 949 
\bibitem[Hodapp et~al.(2009)]{Hod09} Hodapp, K.~W., Iserlohe, C., Stecklum, B., \& Krabbe, A.\ 2009, \apjl, 701, L100 
\bibitem[Horch et~al.(2001)]{Hor01} Horch, E., Ninkov, Z., \& Franz, O.~G.\ 2001, \aj, 121, 1583 
\bibitem[Horch et~al.(2004)]{Hor04} Horch, E.~P., Meyer, R.~D., \& van Altena, W.~F.\ 2004, \aj, 127, 1727 
\bibitem[Hormuth et~al.(2008)]{Hor08} Hormuth, F., Hippler, S., Brandner, W., Wagner, K., \& Henning, T.\ 2008, \procspie, 7014,  
\bibitem[Hummel et~al.(2013)]{Hum13} Hummel, C.~A., Zavala, R.~T., \& Sanborn, J.\ 2013, Central European Astrophysical Bulletin, 37, 127 
\bibitem[Johnson et~al.(1966)]{Joh66} Johnson, H.~L., Mitchell, R.~I., Iriarte, B., \& Wisniewski, W.~Z.\ 1966, Communications of the Lunar and Planetary Laboratory, 4, 99 
\bibitem[Kaufer et~al.(1999)]{Kau99} Kaufer, A., Stahl, O., Tubbesing, S., et~al.\ 1999, The Messenger, 95, 8 
\bibitem[Landstreet \& Borra(1978)]{Lan78} Landstreet, J.~D., \& Borra, E.~F.\ 1978, \apjl, 224, L5 
\bibitem[Lee(1968)]{Lee68} Lee, T.~A.\ 1968, \apj, 152, 913 
\bibitem[Lutz \& Kelker(1973)]{Lut73} Lutz, T.~E., \& Kelker, D.~H.\ 1973, \pasp, 85, 573 
\bibitem[Ma{\'{\i}}z Apell{\'a}niz(2001)]{Mai01} Ma{\'{\i}}z-Apell{\'a}niz, J.\ 2001, \aj, 121, 2737 
\bibitem[Ma{\'{\i}}z Apell{\'a}niz(2004)]{Mai04} Ma{\'{\i}}z-Apell{\'a}niz, J.\ 2004, \pasp, 116, 859 
\bibitem[Ma{\'{\i}}z Apell{\'a}niz(2005a)]{Mai05a} Ma{\'{\i}}z-Apell{\'a}niz, J.\ 2005a, \pasp, 117, 615 
\bibitem[Ma{\'{\i}}z Apell{\'a}niz(2005b)]{Mai05b} Ma{\'{\i}}z Apell{\'a}niz, J.\ 2005b, The Three-Dimensional Universe with Gaia, 576, 179 
\bibitem[Ma{\'{\i}}z Apell{\'a}niz(2006)]{Mai06} Ma{\'{\i}}z Apell{\'a}niz, J.\ 2006, \aj, 131, 1184 
\bibitem[Ma{\'{\i}}z Apell{\'a}niz(2007)]{Mai07} Ma{\'{\i}}z Apell{\'a}niz, J.\ 2007, The Future of Photometric, Spectrophotometric and Polarimetric Standardization, 364, 227 
\bibitem[Ma{\'{\i}}z Apell{\'a}niz(2010)]{Mai10} Ma{\'{\i}}z Apell{\'a}niz, J.\ 2010, \aap, 518, A1 
\bibitem[Ma{\'{\i}}z Apell{\'a}niz(2013a)]{Mai13a} Ma{\'{\i}}z Apell{\'a}niz, J.\ 2013a, Highlights of Spanish Astrophysics VII, 657 
\bibitem[Ma{\'{\i}}z Apell{\'a}niz(2013b)]{Mai13b} Ma{\'{\i}}z Apell{\'a}niz, J.\ 2013b, Highlights of Spanish Astrophysics VII, 583 
\bibitem[Ma{\'{\i}}z Apell{\'a}niz et~al.(2004)]{Mai04+} Ma{\'{\i}}z-Apell{\'a}niz, J., Walborn, N.~R., Galu{\'e}, H.~{\'A}., \& Wei, L.~H.\ 2004, \apjs, 151, 103 
\bibitem[Ma{\'{\i}}z Apell{\'a}niz et~al.(2012)]{Mai12} Ma{\'{\i}}z Apell{\'a}niz, J., Pellerin, A., Barb{\'a}, R.~H., et~al.\ 2012, Proceedings of a Scientific Meeting in Honor of Anthony F.~J.~Moffat, 465, 484 
\bibitem[Ma{\'{\i}}z Apell{\'a}niz et~al.(2014)]{Mai14} Ma{\'{\i}}z Apell{\'a}niz, J., Evans, C.~J., Barb{\'a}, R.~H., et~al.\ 2014, \aap, 564, A63 
\bibitem[Mayne \& Naylor(2008)]{May08} Mayne, N.~J., \& Naylor, T.\ 2008, \mnras, 386, 261 
\bibitem[Miczaika(1950)]{Mic50} Miczaika, G.~R.\ 1950, \apj, 111, 443 
\bibitem[Morrell \& Levato(1991)]{Mor91} Morrell, N., \& Levato, H.\ 1991, \apjs, 75, 965 
\bibitem[Najarro et~al.(2011)]{Naj11} Najarro, F., Hanson, M.~M., \& Puls, J.\ 2011, \aap, 535, A32 
\bibitem[Naylor(2009)]{Nay09} Naylor, T.\ 2009, \mnras, 399, 432 
\bibitem[Nieva \& Sim{\'o}n-D{\'{\i}}az(2011)]{Nie11} Nieva, M.-F., \& Sim{\'o}n-D{\'{\i}}az, S.\ 2011, \aap, 532, AA2 
\bibitem[Ogura \& Sugitani(1998)]{Ogu98} Ogura, K., \& Sugitani, K.\ 1998, \pasa, 15, 91 
\bibitem[Perryman et~al.(1997)]{Per97} Perryman, M.~A.~C., Lindegren, L., Kovalevsky, J., et~al.\ 1997, \aap, 323, L49 
\bibitem[Pety et~al.(2005)]{Pet05} Pety, J., Teyssier, D., Foss{\'e}, D., et~al.\ 2005, \aap, 435, 885 
\bibitem[Pound et~al.(2003)]{Pou03} Pound, M.~W., Reipurth, B., \& Bally, J.\ 2003, \aj, 125, 2108 
\bibitem[Raskin et~al.(2011)]{Ras11} Raskin, G., van Winckel, H., Hensberge, H., et~al.\ 2011, \aap, 526, A69 
\bibitem[Repolust et~al.(2004)]{Rep04} Repolust, T., Puls, J., \& Herrero, A.\ 2004, \aap, 415, 349 
\bibitem[Rimmer et~al.(2012)]{Rim12} Rimmer, P.~B., Herbst, E., Morata, O., \& Roueff, E.\ 2012, \aap, 537, A7 
\bibitem[Rivilla et al.(2013)]{Riv13} Rivilla, V.~M., Mart{\'{\i}}n-Pintado, J., Jim{\'e}nez-Serra, I., \& Rodr{\'{\i}}guez-Franco, A.\ 2013, \aap, 554, A48 
\bibitem[Sacco et~al.(2008)]{Sac08} Sacco, G.~G., Franciosini, E., Randich, S., \& Pallavicini, R.\ 2008, \aap, 488, 167 
\bibitem[Salpeter(1955)]{Sal55} Salpeter, E.~E.\ 1955, \apj, 121, 161 
\bibitem[Sanz-Forcada et~al.(2004)]{San04} Sanz-Forcada, J., Franciosini, E., \& Pallavicini, R.\ 2004, \aap, 421, 715 
\bibitem[Schaefer(2013)]{Sch13} Schaefer, G.~H.\ 2013, EAS Publications Series, 64, 181 
\bibitem[Schneider et al.(2014a)]{Sch14a} Schneider, F.~R.~N., Izzard, R.~G., de Mink, S.~E., et al.\ 2014a, \apj, 780, 117 
\bibitem[Schneider et al.(2014b)]{Sch14b} Schneider, F.~R.~N., Langer, N., de Koter, A., et al.\ 2014b, \aap, 570, AA66 
\bibitem[Sherry et~al.(2004)]{She04} Sherry, W.~H., Walter, F.~M., \& Wolk, S.~J.\ 2004, \aj, 128, 2316 
\bibitem[Sherry et~al.(2008)]{She08} Sherry, W.~H., Walter, F.~M., Wolk, S.~J., \& Adams, N.~R.\ 2008, \aj, 135, 1616 
\bibitem[Sim{\'o}n-D{\'{\i}}az(2010)]{Sim10} Sim{\'o}n-D{\'{\i}}az, S.\ 2010, \aap, 510, AA22 
\bibitem[Sim{\'o}n-D{\'{\i}}az \& Herrero(2014)]{Sim14} Sim{\'o}n-D{\'{\i}}az, S., \& Herrero, A.\ 2014, \aap, 562, A135 
\bibitem[Sim{\'o}n-D{\'{\i}}az et~al.(2011a)]{Sim11a} Sim{\'o}n-D{\'{\i}}az, S., Caballero, J.~A., \& Lorenzo, J.\ 2011a, \apj, 742, 55 
\bibitem[Sim{\'o}n-D{\'{\i}}az et~al.(2011b)]{Sim11b} Sim{\'o}n-D{\'{\i}}az, S., Castro, N., Garcia, M., Herrero, A., \& Markova, N.\ 2011b, Bulletin de la Societe Royale des Sciences de Liege, 80, 514 
\bibitem[Sim{\'o}n-D{\'{\i}}az et~al.(2011c)]{Sim11c} Sim{\'o}n-D{\'{\i}}az, S., Castro, N., Herrero, A., et~al.\ 2011c, Journal of Physics Conference Series, 328, 012021 
\bibitem[Sim{\'o}n-D{\'{\i}}az et~al.(2011d)]{Sim11d} Sim{\'o}n-D{\'{\i}}az, S., Garcia, M., Herrero, A., Ma{\'{\i}}z Apell{\'a}niz, J., \& Negueruela, I.\ 2011d, Stellar Clusters \& Associations: A RIA Workshop on Gaia, 255 
\bibitem[Skinner et~al.(2008)]{Ski08} Skinner, S.~L., Sokal, K.~R., Cohen, D.~H., et~al.\ 2008, \apj, 683, 796 
\bibitem[Stickland \& Lloyd(2001)]{Sti01} Stickland, D.~J., \& Lloyd, C.\ 2001, The Observatory, 121, 1 
\bibitem[Telting et~al.(2014)]{Tel14} Telting, J.~H., \'Avila, G., Buchhave, L., et~al.\ 2014, AN, 335, 41 
\bibitem[ten Brummelaar et~al.(2000)]{ten00} ten Brummelaar, T., Mason, B.~D., McAlister, H.~A., et~al.\ 2000, \aj, 119, 2403 
\bibitem[Tokovinin(2008)]{Tok08} Tokovinin, A.\ 2008, \mnras, 389, 925 
\bibitem[Townsend et~al.(2013)]{Tow13} Townsend, R.~H.~D., Rivinius, T., Rowe, J.~F., et~al.\ 2013, \apj, 769, 33 
\bibitem[Tull(1998)]{Tul98} Tull, R.~G.\ 1998, \procspie, 3355, 387 
\bibitem[Turner et~al.(2008)]{Tur08} Turner, N.~H., ten Brummelaar, T.~A., Roberts, L.~C., et~al.\ 2008, \aj, 136, 554 
\bibitem[van Leeuwen(2007)]{van07} van Leeuwen, F.\ 2007, \aap, 474, 653 
\bibitem[van Loon \& Oliveira(2003)]{van03} van Loon, J.~T., \& Oliveira, J.~M.\ 2003, \aap, 405, L33 
\bibitem[van Bever \& Vanbeveren(1998)]{van98} van Bever, J., \& Vanbeveren, D.\ 1998, \aap, 334, 21 
\bibitem[Vogt(1976)]{Vog76} Vogt, N.\ 1976, \aap, 53, 9 
\bibitem[Walborn(1974)]{Wal74} Walborn, N.~R.\ 1974, \apjl, 191, L95 
\bibitem[Walter et~al.(2008)]{Wal08} Walter, F. M., Sherry, W. H., Wolk, S. J., Adams, N. R. 2008, {\em Handbook of Star Forming Regions, Volume I: The Northern Sky} ASP Monograph Publications, vol. 4. Ed. by Bo~Reipurth, p.~732
\bibitem[Ward-Thompson et~al.(2006)]{War06} Ward-Thompson, D., Nutter, D., Bontemps, S., Whitworth, A., \& Attwood, R.\ 2006, \mnras, 369, 1201 
\bibitem[Whitworth et~al.(2007)]{Whi07} Whitworth, A., Bate, M.~R., Nordlund, {\AA}., Reipurth, B., 
\& Zinnecker, H.\ 2007, Protostars and Planets V, 459 
\bibitem[Wolk(1996)]{Wol96} Wolk, S. J. 1996, Ph.D. thesis, State University of New York, NY, USA
\bibitem[Zapatero Osorio et~al.(2002)]{Zap02} Zapatero Osorio, M.~R., B{\'e}jar, V.~J.~S., Pavlenko, Y., et~al.\ 2002, \aap, 384, 937 
\bibitem[Zinnecker \& Yorke(2007)]{Zin07} Zinnecker, H., \& Yorke, H.~W.\ 2007, \araa, 45, 481 
\end{thebibliography}
\end{document}